\documentclass[bib]{statapress}

\usepackage[crop,newcenter,frame]{pagedims}

\usepackage{sj}

\usepackage{epsfig}

\usepackage{stata}

\usepackage{amsmath}

\usepackage{shadow}

\sjsetissue{$vv$}{$ii$}{$mm$}{$yyyy$}

\usepackage{threeparttable}
\usepackage{tabularx}
\usepackage{comment}

\usepackage{lmodern}

\usepackage[T1]{fontenc}

\usepackage{textcomp}

\usepackage{color}

\usepackage{soul}

\begin{document}
\inserttype[st0001]{article}
\author{J. Ditzen, Y. Karavias and J. Westerlund}{%
	Jan Ditzen\\Free University of Bozen-Bolzano\\Bozen/Italy\\jan.ditzen@unibz.it
	\and
	Yiannis Karavias\\Brunel University of London\\London/UK\\i.karavias@bham.ac.uk
	\and \\
	Joakim Westerlund\\Lund University\\Lund/Sweden\\joakim.westerlund@nek.lu.se\\and\\Deakin University\\Melbourne/Australia\\j.westerlund@deakin.edu.au
}
\title[Structural Breaks in Stata]{Testing and Estimating Structural Breaks in Time Series and Panel Data in Stata}
\maketitle
\thispagestyle{empty}
\begin{abstract}
Identifying structural change is a crucial step in analysis of time series and panel data. The longer the time span, the higher the likelihood that the model parameters have changed as a result of major disruptive events, such as the 2007--2008 financial crisis and the 2020 COVID--19 outbreak. Detecting the existence of breaks, and dating them is therefore necessary, not only for estimation purposes but also for understanding drivers of change and their effect on relationships. This article introduces a new community contributed command called \texttt{xtbreak}, which provides researchers with a complete toolbox for analysing multiple structural breaks in time series and panel data. \texttt{xtbreak} can detect the existence of breaks, determine their number and location, and provide break date confidence intervals. The new command is used to explore changes in the relationship between COVID--19 cases and deaths in the US, using both aggregate and state level data, and in the relationship between approval ratings and consumer confidence, using a panel of eight countries.   \\

\keywords{Structural breaks; Change points; Time series data; Panel data; Interactive Fixed Effects; Cross-section Dependence; \stcmd{xtbreak}.} \\

\end{abstract}

\section{Introduction}

In economics and elsewhere linear relationships between dependent and explanatory variables are at the core of interest. To investigate such relationships, observations over time for one or more cross-sectional units such as firms, individuals, or countries are collected and are subsequently used in estimating the coefficients of regression models. A key assumption here is that the coefficients do not change over time. This assumption is unlikely to hold, especially for longer periods of time, because of major disruptive events, such as financial crises. Parameter instability can have a detrimental impact on estimation and inference, and can lead to costly errors in decision-making. The times in which the parameters change are called ``change points'' in the statistics literature and ``structural breaks'' in economics. As both terms are synonymous to each other, in the following we will use the latter term or just ``breaks''.

The aim of this paper is to propose a new community contributed Stata package called \texttt{xtbreak}.\footnote{Updates will be continuously provided on our GitHub page: \href{https://janditzen.github.io/xtbreak/}{GitHub}.} The package implements the methods developed by \cite{BaiPerron1998} for the case of pure time series, and \cite{DKW2021} in the case of panel data.\footnote{The \cite{DKW2021} study develops the methods and asymptotic theory for the analysis of panel data with multiple structural breaks and interactive fixed effects. For one break, the results coincide with the earlier work by \cite{KNW2021}.}

\texttt{xtbreak} provides researchers with a complete toolbox for analysing multiple structural breaks in time series and panel data. It can detect and date an unknown number of breaks at unknown break dates. The toolbox is based on asymptotically valid tests for the presence of breaks, a consistent break date estimator, and a break date confidence interval with correct asymptotic coverage. In fact, \texttt{xtbreak} includes no less than three tests; (i) a test of no structural breaks against the alternative of a specific number of breaks, (ii) a test the null hypothesis of no structural breaks against the alternative of an unknown number of structural breaks, and (iii) a test of the null of $s$ breaks against the alternative of $s+1$ breaks. The package also includes an algorithm that employs the last test consecutively in order to estimate the true number of breaks. The tested break dates can be unknown or user-defined, as when researchers have additional information and wish to examine whether there was a break in a specific point in time. Once the presence of breaks has been tested and confirmed, \texttt{xtbreak} estimates the locations of the breaks and provides the associated confidence intervals.

A large number of breaks does not translate into heavy computational burden, as \texttt{xtbreak} implements an efficient dynamic programming method described in \cite{Bai2003}, which ensures that there are $ O(T^2) $ computations even with more than two breaks, where $T$ is the number of time series observations.

\texttt{xtbreak} can deal with models of ``pure'' or ``partial'' structural breaks. A pure structural breaks model is one in which the coefficients of all explanatory variables change, while in a partial structural breaks model only a subset of the coefficients change.

\texttt{xtbreak} is applicable under very general error conditions. For time series data the only requirement is that there are no unit roots in the errors. In case of panel data, units can be independent, or cross-sectionally dependent where cross-sectional dependence takes an ``interactive fixed effects'', or ``common factor'', structure. Regressors can load on the same set of factors as the errors, which means that regressors may be endogenous -  although no instrumental variables are necessary. The errors can also be serially correlated and heteroskedastic, but not non-stationary.

The works of \cite{BaiPerron1998}, and \cite{DKW2021} draw on the economics literature. However, structural breaks are not confined to economics but happen also in other fields of research, including engineering, epidemiology, climatology, and medicine. \texttt{xtbreak} is therefore widely applicable. To showcase this width, we consider two examples drawn from the areas of epidemiology and political economy. First, we consider the epidemiological relationship between COVID--19 cases and deaths. Using both aggregate country and disaggregated state level US data, we find evidence of multiple breaks. In particular, we find that an increase in the number of COVID--19 cases lead to more deaths in the beginning of the pandemic than in later waves. Secondly, we examine if there are breaks in the relationship between consumer confidence and the approval ratings of country leaders. Using a panel of eight countries observed over a long period of time we find that there is great cross-country heterogeneity in terms of the number and locations of breaks.  

The remainder of the paper is organized as follows: Section 2 presents the model that we will be considering. We focus on the panel case, which in most regards includes the pure time series setup as a special case. Important differences are brought up and discussed. Sections 3, 4 and 5 present the hypothesis tests, the break date estimation procedure, and the \texttt{xtbreak} command, respectively. Sections 6 and 7 contain the empirical analyses of the COVID--19 and leader approval ratings data, respectively. Section 8 concludes the paper. Section 9 presents instructions on installing \texttt{xtbreak}, and Sections 10 and 11 contain acknowledgements and references, respectively.

\section{Model discussion}

We consider the following model with $N$ units, $T$ periods and $s$ structural breaks:
\begin{align}
y_{i,t} =x_{i,t}'\beta+w_{i,t}'\delta_j+e_{i,t} ,\label{eq:y}
\end{align}
where $t=T_{j-1},...,T_j$ and $j=1,...,s+1$ with $T_0=0$ and $T_{s+1}=T$. Hence, there are $s$ breaks, or $s+1$ regimes with regime $j$ covering the observations $T_{j-1},...,T_j$. In order emphasise the break structure, we can write \eqref{eq:y} regime-wise;
\begin{align*}
y_{i,t} &=x_{i,t}'\beta+w_{i,t}'\delta_1+e_{i,t} ~ \text{for} ~ t = T_0,...,T_1,\\
y_{i,t} &=x_{i,t}'\beta+w_{i,t}'\delta_2+e_{i,t} ~ \text{for} ~ t = T_1,...,T_2,\\
& \vdots \\
y_{i,t} &=x_{i,t}'\beta+w_{i,t}'\delta_{s+1}+e_{i,t} ~ \text{for} ~ t = T_{s},...,T_{s+1}.
\end{align*}
For $ N=1 $, this is a time series model, while for $ N>1 $, it is a panel data model. The dependent variable $y_{i,t}$ and the regression error $e_{i,t}$ are scalars, while $x_{i,t}$ and $w_{i,t}$ are $p\times 1$ and $q\times 1$ vectors, respectively, of regressors. The coefficients of the regressors in $x_{i,t}$ are unaffected by the breaks, while those of $w_{i,t}$ are affected by the breaks. It is possible that all independent variables break, in which case $x_{i,t}'\beta$ is defined to be zero. In the panel case, the break dates are common for all units. This is a very common assumption that is reasonable in settings where the frequency of the data is not high. Let $\mathcal{T}_s=\{T_1,...,T_s\}$ be a collection of $s$ break dates such that $ T_j=\lfloor \lambda_j T \rfloor $, where $\lambda_0 = 0 < \lambda_1 < ... <\lambda_{s} < \lambda_{s+1} = 1$. By specifying the breaks in this way we ensure that they are distinct from one another and that they are bounded away from the beginning and end of the sample. This is important because we need to be able to estimate the model within each regime.

In case the data has a panel structure ($N > 1$), we allow for unobserved heterogeneity in the form of interactive fixed effects:
\begin{equation}
	e_{i,t}=f_t'\gamma_i+\varepsilon_{i,t},	\label{eq:e}
\end{equation}
where $ f_t $ is an $ m\times 1 $ vector of factors and $ \gamma_i $ is a conformable vector of factor loadings.\footnote{We use the term ``unobserved heterogeneity'', although $f_t$ might be known. Later on we elaborate on this.} The fact that $f_t$ is common to all cross-sectional units $i$ means that the regression errors can be strongly cross-sectionally correlated. This specification is very general and nests the usual one-way and two-way fixed effects models as special cases. Both $f_t$ and $\varepsilon_{i,t}$ may be weakly serially correlated, but they cannot be nonstationary. They can also not be correlated with each other, and $\varepsilon_{i,t}$ cannot be cross-sectionally correlated.\footnote{\cite{BaiPerron1998} clarify that in the presence of nonstationary regressors the break date consistency and rate of convergence remains, but not the limiting distributions. \cite{DKW2021} conjecture that the factors can be non-stationary, without any differences to the methodology.} This last condition ensures that any cross-section dependence in $e_{i,t}$ originates with $f_{t}$.

Typically there is a lot of cross-sectional co-movement not only in the regression errors but also in the regressors. To account for this, we assume that $ x_{i,t} $ and $ w_{i,t} $ are generated in the following way:
\begin{align}
	x_{it}=\Gamma_{x,i}'f_t+u_{x,i,t},\label{eq:x}\\
	w_{it}=\Gamma_{w,i}'f_t+u_{w,i,t},\label{eq:w}
\end{align}
where $ \Gamma_{x,i} $ and $ \Gamma_{w,i} $ are $ p\times m $ and $ q\times m $ matrices, respectively, of factor loadings, while $ u_{x,i,t}$ and $u_{w,i,t} $ are $ p\times 1 $ and $ q\times 1 $ vectors, respectively, of idiosyncratic errors that are independent of all the other random elements of the model. The model described by \eqref{eq:y}--\eqref{eq:w} above is the same as the one considered by \cite{DKW2021}.\footnote{The factor-in-regressors condition rules out lags of the dependent variable as regressors in the panel case. However, this is not really a restriction on the data generating process, as general forms of serial correlation are still permitted through both factors and idiosyncratic errors. Therefore, one can estimate a static model without the lagged dependent variable, which is left as serial correlation in the errors. The main limitations of doing this are: (i) inefficiency since we do not account for the serial correlation in the estimation, and (ii) loss of interpretation in some models where the coefficient of the lagged dependent variable is the parameter of interest. Limitation (i) may not be important in moderate or large panels. Lagged dependent variables are not ruled out in the time series case (see \citealp{BaiPerron1998}).}

The fact that $ x_{i,t} $ and $ w_{i,t} $ are allowed to load on the same set of factors as $ e_{i,t} $ means that they can be endogenous. This type of endogeneity through unobserved heterogeneity is standard in panel data. Here we are considering interactive fixed effects, but the idea is the same in the fixed effects case; the effects sitting in errors might also be in the regressors, which means that they have to be removed prior to estimation. In the fixed effects case, one augments \eqref{eq:y} by dummy variables, which is tantamount to transforming the variables into deviations from means. Here a more elaborate augmentation approach is needed, as to be expected since interactive effects are more general than fixed effects. Had the factors been known, which would be the case if the unobserved heterogeneity is made up of known deterministic terms for example, we would have estimated
\begin{align}
y_{i,t} =x_{i,t}'\beta+w_{i,t}'\delta_j+f_t'\gamma_i+\varepsilon_{i,t}\label{eq:fknown}
\end{align}
by ordinary least squares (OLS). This is possible because with $f_t$ as a regressor the regression error is no longer given by $e_{i,t}$ but by $\varepsilon_{i,t}$, which is independent of the regressors. If, on the other hand, $f_t$ is not known, then we need a good proxy to use in its stead.\footnote{Take the so-called ``wage curve'' model, which relates worker's wages to the rate of unemployment, and is the most common motivating example in the literature (see, for example, Bai, 2009). Here the factor loadings may represent workers' unobservable skills, such as innate ability, perseverance, and motivation, and the factors would represent the price of these skills, which are not necessarily
constant over time.} \cite{DKW2021} use $\bar x_t = N^{-1}\sum_{i=1}^N x_{i,t}$ and $\bar w_t = N^{-1}\sum_{i=1}^N w_{i,t}$, and therefore so do we.\footnote{The intuition for why argumentation by $\bar x_{t}$ is needed is simple. We begin by noting that by \eqref{eq:x}, $\bar x_{t}=\bar \Gamma_{x}'f_t+\bar u_{x,t}$, where $\bar \Gamma_{x}$ and $\bar u_{x,t}$ are the cross-sectional averages of $\Gamma_{x,i}$ and $u_{x,i,t}$, respectively. Hence, provided that $u_{x,i,t}$ is mean zero and independent across $i$, by a central limit law, $\bar x_{t}\to_p\bar \Gamma_{x}'f_t$ as $N\to\infty$, where ``$\to_p$'' signify convergence in probability. We say that $\bar x_{t}$ is ``rotationally consistent'' for $f_t$, which is enough if the purpose is to control for $f_t$. The intuition for why $\bar w_{t}$ is need is analogous.} The appropriately augmented version of \eqref{eq:y} is therefore given by
\begin{align}
y_{i,t} =x_{i,t}'\beta+w_{i,t}'\delta_j+\bar x_{t}'a_i+\bar w_{t}'b_{i,j}+\text{error}. \label{eq:est}
\end{align}
Because asymptotically observing the cross-sectional averages is just as good as observing the true factors, the regressors in \eqref{eq:est} are asymptotically exogenous. This means that the estimation can be carried out using OLS. This is the same idea as in the ``common correlated effects'' (CCE) estimator of \citep{Pesaran2006}, with the difference that here we do not include the cross-sectional average of $y_{i,t}$ as a regressor in \eqref{eq:est} (see \citealp{KNW2021}, for a discussion). If $f_t$ is neither completely known, nor completely unknown, as is usually the case in practice, then the cross-sectional averages will take care of the unknown factors and the known factors can be added to \eqref{eq:est} as additional regressors.\footnote{The cross-sectional averages can in principle capture all factors, regardless of whether they are observed (known) or not. An important condition for this to work is, however, that the number of cross-section averages is not fewer than the number of factors they replace. Because of this it is a good idea to treat known factors as additional regressors in \eqref{eq:est}, as it makes it possible to estimate more unknown factors.}

If $ N=1 $, such that \eqref{eq:y} is a time series model, then by definition there is no cross-sectional variation that we can exploit to estimate unknown factors. Hence, in this case $f_t$ must be known, and hence the model to be estimated is given by \eqref{eq:fknown}, which is then the same as in \cite{BaiPerron1998}.

\section{The toolbox}

\subsection{Hypotheses testing}\label{section::test}

This section presents the first set of tools, which are necessary for establishing that one or more structural breaks have happened and for determining their number, $s$. In particular, we consider tests of three hypotheses, labelled ``(1)''--``(3)'', and a sequential test to determine $s$. We begin by stating the hypotheses of interest.

\begin{enumerate}
	\item[(1)] $H_0$: no breaks versus $H_1$: $s$ breaks, where the number of breaks under $H_1$, $s$, is specified by the researcher.

	\item[(2)] $H_0$: no breaks versus $H_1$: $1\leq s\leq s_{max}$ breaks, where the maximum number of breaks under $H_1$, $s_{max}$, is specified by the researcher.

	\item[(3)] $H_0$: $s$ breaks versus $H_1$: $s+1$ breaks, where $s$ is specified by the researcher.
\end{enumerate}

To test hypotheses (1)--(3) we employ a number of test statistics. The time series versions of these tests have appeared in \cite{BaiPerron1998}, while the panel versions have appeared in \cite{DKW2021}.

\subsubsection*{Hypothesis (1)}

If the dates of the breaks are known, the test that we are going to consider for hypothesis (1) is simply a Chow test. Let us therefore denote by $F(\mathcal{T}_s)$ the $F$-statistic for testing the null of no breaks versus the alternative of $s$ known breaks at dates $\mathcal{T}_s$, which in the time series case is based on \eqref{eq:fknown}, while in the panel case it is based on \eqref{eq:est}. Appropriate critical values can be taken from the $F$ distribution with $s$ numerator degrees of freedom and $ N(T-p-(s+1)q)-p-(s+1)q $ denominator degrees of freedom.

If $\mathcal{T}_s$ is unknown, which is most likely the case in practice, then the following supremum statistic can be used:
\begin{equation}
\mathrm{sup} F(s) =\sup_{\mathcal{T}_s\in \mathcal{T}_{s,\epsilon}} F(\mathcal{T}_s). \label{eq:supwald}
\end{equation}
Here
\begin{equation}
\mathcal{T}_{s,\epsilon} = \{(T_1,...,T_s):T_{j+1}-T_{j} \geq \epsilon T, T_1\geq \epsilon T, T_s\leq (1-\epsilon) T \},
\end{equation}
is the set of permissible break dates with $\epsilon$ being a user-defined trimming parameter. By setting $\epsilon \in (0,1)$ we ensure that the breaks considered in the test are distinct and bounded away from the sample endpoints, as assumed.

\subsubsection*{Hypothesis (2)}

Hypothesis (2) can be tested using the following double maximum statistic:
\begin{equation}
\mathrm{WDmax}F(s_{max})= \max_{1\leq s\leq s_{max}} \frac{c_{\alpha,1}}{c_{\alpha,s}} \mathrm{sup}F(s),
\label{eq:doublemax}
\end{equation}
where $c_{\alpha,s}$ is the critical value of $\mathrm{sup}F(s)$ at significance level $\alpha$ and $s$ breaks. The weighting by $c_{\alpha,1}/c_{\alpha,s}$ here ensures that the marginal $p$-values of the weighted supremum statistics are all equal. This counterweights the decrease in the marginal $p$-value of $\mathrm{sup}F(s)$ that comes from increasing $s$, and the resulting loss of power when $s$ is large. The test statistic with weights $c_{\alpha,1}/c_{\alpha,s} = 1$ for all $s$ is called $\mathrm{UDmax}F(s_{max})$.

\subsubsection*{Hypothesis (3)}

In Section \ref{section::breakestim}, we describe a procedure for how to estimate the break dates. Let $\hat{\mathcal{T}}_s=\{\hat T_1,...,\hat T_s\}$ be the set of estimated breaks obtained based on that procedure. For the test of hypothesis (3) we use the following statistic:
\begin{equation}
F(s+1|s)= \sup_{1\leq j \leq s+1} \sup_{\tau \in \hat{\mathcal{T}}_{j,\epsilon}} F(\tau |\hat{\mathcal{T}}_s).
\label{eq:seqf}
\end{equation}
where $\hat{\mathcal{T}}_{s}$ contains estimates of the $s$ break stipulated under the hull hypothesis, $\tau$ is the additional $(s+1)$-th break under the alternative, and
\begin{equation}
\hat{\mathcal{T}}_{j,\epsilon} = \{\tau : \hat T_{j-1}+ (\hat T_j - \hat T_{j-1})\epsilon \leq \tau \leq  \hat T_{j}- (\hat T_j-\hat T_{j-1})\epsilon, \hat T_0=0, \hat T_{s+1}=1 \}
\end{equation}
is the set of permissible breaks in between the estimated $(j-1)$-th and $j$-th breaks. Hence, $F(s+1|s)$ is testing the null of $s$ breaks versus the alternative that there is an additional break somewhere within the regimes stipulated under the null. Finally, $F(\tau|\hat{\mathcal{T}}_s)$ is the $F$-statistic based on taking the estimated break dates in $\hat{\mathcal{T}}_s$ as given and testing for one additional break at $\tau$.

The $F(s+1|s)$ test can be applied sequentially to estimate the number of breaks. In this case, we start by testing the null of no breaks against the alternative of a single break using $F(1|0)$. If the null is accepted, we set $\hat s = 0$ and terminate the procedure. If, however, the null is rejected, we estimate the breakpoint, denoted $\hat T_1$, and split the sample in two at $\hat T_1$. We then test for the presence of a break in each of the two subsamples using $F(2|1)$. If no breaks are found, we set $\hat s = 1$ and stop, whereas if breaks are detected, we estimate their location and split the sample again. This process continues until the test fails to reject.

The asymptotic distributions of the above tests in the pure time series and panel cases can be found in \cite{BaiPerron1998}, and \cite{DKW2021}, respectively. Because the distributions are the same, so are the critical values. \texttt{xtbreak} therefore uses the critical values of \cite{BaiPerron1998}, and \cite{Bai2003}, which are applicable for $ \epsilon\in\{0.05,0.1,0.15,0.2,0.25\} $. In theory, the validity of the critical values requires $ T\to\infty $ in the time series case and $ N,\,T\to\infty $ with $ T/N\to 0$ in the panel case, which in practice means that $T$ should be ``large'' in both cases, and that $N$ should be even larger in the panel case. For some Monte Carlo evidence on the accuracy of these predictions, we make reference to \cite{Bai2003}, and \cite{DKW2021}.

\section{Break date estimation}\label{section::breakestim}

The previous section was focused on testing for the existence of breaks, and on determining their number. As soon as the number of breaks is detected, interest turns to their location, as it is the dates of the breaks that researchers use to identify the underlying cause of breaks.

The standard approach in the literature is to estimate breaks by minimizing the sum of squared residuals. \cite{BaiPerron1998}, and \cite{DKW2021} do the same. The break date estimator included in \texttt{xtbreak} is therefore given by
\begin{equation}
\hat{\mathcal{T}}_{s} =\arg \min_{\mathcal{T}_{s} \in \mathcal{T}_{s,\varepsilon}} SSR(\mathcal{T}_{s}) ,\label{eq:breakdateestim}
\end{equation}
where $SSR(\mathcal{T}_s)$ is the sum of squared residuals based on $s$ breaks. In the time series case the residuals are taken from \eqref{eq:fknown}, whereas in the panel case they are taken from \eqref{eq:est}. If $s$ is ``small'', the minimization can be done by grid search. If, however, $s$ is ``large'', then grid search, which requires $O(T^s)$ OLS operations, becomes computationally very costly and possibly even infeasible. In such cases, the efficient dynamic programming algorithms of \cite{BaiPerron1998,Bai2003}, and \cite{DKW2021}, which limit the number of operations to $O(T^2)$ for any $s$, can be used.

Once $\hat{\mathcal{T}}_{s}$ has been obtained, confidence intervals for each estimated break date can be constructed using the formulas given in \cite{BaiPerron1998}, and \cite{DKW2021}.

\section{The \texttt{xtbreak} command}
\label{sec:syntax}
\subsection{Syntax}

\subsubsection{Automatic estimation of number of breaks and breakdates}

\begin{stsyntax}
	xtbreak \depvar\ \optional{{\it indepvars}} \optif\ \optional{, \textit{options1} \textit{options2} \textit{options3} \textit{options5} \textit{options6} }
\end{stsyntax}

tests for breaks via hypothesis (2) and estimates the number of breaks and breakdates with no prior knowledge on number and location of breaks. Estimation of the number of breaks is based on the sequential test of hypothesis (3).

\subsubsection{Testing for known structural breaks}

\noindent 
\begin{stsyntax}
	xtbreak test 	\depvar\ \optional{{\it indepvars}} \optif\ ,
		\underbar{breakp}oints(numlist|datelist \optional{,index|fmt(\ststring)}) \optional{\textit{options1} \textit{options5} }
\end{stsyntax}

implements hypothesis (1): testing for breaks if the break dates are known.

\subsubsection{Testing for unknown structural breaks}

\begin{stsyntax}
	xtbreak test 	\depvar\ \optional{{\it indepvars}} \optif\ \optional{,
		\underline{h}ypothesis(1|2|3) breaks(real) \textit{options1} \textit{options2} \textit{options3} \textit{options4} \textit{options5} }
\end{stsyntax}

implements hypotheses (1)-(3): testing for breaks if the break dates are unknown. 

The default is hypothesis (3) and option \texttt{sequential}.

\subsubsection{Estimation of breakdates}

\begin{stsyntax}
	xtbreak estimate 	\depvar\ \optional{{\it indepvars}} \optif\ \optional{, breaks(real)  showindex \textit{options1} \textit{options2}  \textit{options5}}
\end{stsyntax}

estimates breakdates for a given number of breaks.

\subsubsection{Updating \texttt{xtbreak}}

\begin{stsyntax}
	xtbreak, update
\end{stsyntax}

obtains the latest version from GitHub.

\subsubsection{Specific Options}

\noindent
\textit{options1} are general options and apply to \texttt{xtbreak} in general:

\indent\texttt{
\underbar{breakcons}tant
\underbar{nocons}tant
\underbar{nobreakvar}iables(varlist1)
vce(type)}

\indent\texttt{\underbar{inv}erter(speed|precision|chol|p|lu) python}

\noindent
\textit{options2} are specific for unknown break dates:

\indent\texttt{
\underbar{trim}ming(real)
}

\noindent
\textit{options3} are specific for unknown breakdates and hypothesis (2):

\indent\texttt{
wdmax level(\#)
}

\noindent
\textit{options4} are specific for unknown breakdates and hypothesis (3):

\indent\texttt{
\underbar{seq}uential
}

\noindent
\textit{options5} are panel data specific:

\indent\texttt{
\underbar{breakf}ixedeffects 
\underbar{nof}ixedeffects 
csd 
csa(\varlist) 
\underbar{csano}break(\varlist)}\\
\indent\texttt{\ \underline{kf}actors(\varlist)
\underline{nbkf}actors(\varlist)    
noreweigh
}

\noindent
\textit{options6} apply to the automatic estimation of the number and location of breaks:

\indent\texttt{
skiph2 cvalue(level) strict max(\#)
}

Data must be \tsref{tsset} or \xtref{xtset} before using \texttt{xtbreak}. Time series data must not include gaps. Panel data can be unbalanced. In this case, observations with missing data will not be included in the regressions. {\it depvar}, {\it indepvars} and \varlist\ may contain time-series operators, see \tsref{tsvarlist}.

\subsection{Description of options}

\hangpara
\texttt{xtbreak} detects automatically if $ N=1 $ or $ N>1 $. The user therefore does not have to specify if the data have a time series or a panel structure.

\hangpara
\texttt{\underbar{breakp}oints(numlist|datelist [,index|fmt(format)])} specifies the known breakpoints. Known breakpoints can be set by either the number of the corresponding observations or by the values of the time identifier. If a \textit{numlist} is used, option index is required.  For example, \texttt{breakpoints(10,index)} specifies that the one break occurs at the 10-th observation as ordered by time. \textit{datelist} takes a list of dates. For example, \texttt{breakpoints(2010Q1, fmt(tq))} specifies a break in the first quarter of 2010. The option \texttt{fmt()} specifies the format and is required if a datelist is used. The format set in \texttt{breakpoints()} and the time identifier needs to be the same.

\hangpara
\texttt{breaks(\#)} specifies the number of unknown breaks under the alternative. For hypothesis (2), \texttt{breaks()} can include two values. For example, \texttt{breaks(4 6)} amounts to testing the null of no breaks against the alternative of 4--6 breaks. If only one value specified, then the lower bound of the number of breaks under the alternative is set to 1. If hypothesis (3) is tested, then \texttt{breaks()} defines the number of breaks under the alternative. If hypothesis (3) is tested and \texttt{breaks()} not defined, then option \texttt{sequential} is invoked.

\hangpara
\texttt{showindex} show confidence intervals as index.

\hangpara
\texttt{\underbar{h}ypothesis(1|2|3)} specifies which hypothesis to test. Specify \texttt{h(1)} for hypothesis (1), \texttt{h(2)} for hypothesis (2) and \texttt{h(3)} for hypothesis (3). The default is \texttt{h(3)} in combination with the option \texttt{sequential}.

\hangpara
\texttt{\underbar{breakcons}tant} break in constant. Default is no breaks in the constant term.

\hangpara
\texttt{\underbar{nocons}tant} suppresses constant.

\hangpara
\texttt{\underbar{breakf}ixedeffects} break in individual fixed effects. Default is no breaks in fixed effects.

\hangpara
\texttt{\underbar{nof}ixedeffects} suppresses individual fixed effects.

\hangpara
\texttt{\underbar{nobreakvar}iables(varlist)} defines variables with no structural breaks. \texttt{varlist} can contain time series operators.

\hangpara
\texttt{vce(type)} specifies the covariance matrix estimator. The options are: ssr (homoskedastic errors, the default), hac (heteroskedastic and autocorrelation robust), hc (heteroskedastic robust), wpn (the fixed-$T$ standard errors of \citealp{wpn19}) and np (the non-parametric estimator of \citealp{Pesaran2006}).

\hangpara
\texttt{\underbar{trim}ming(\ststring)} specifies the trimming parameter in percent. The trimming affects the minimal time periods between two breaks. The default is 15\% (0.15). Critical values are available for 5\%, 10\%, 15\%, 20\% and 25\%.

\hangpara
\texttt{wdmax} weights the double maximum test statistic used for testing hypothesis (2). The default is not to use any weights.

\hangpara
\texttt{level(\#)} sets the significance level for critical values for the double maximum test. If a value is chosen for which no critical values exits, \texttt{xtbreak test} will choose the closest level.

\hangpara
\texttt{\underline{seq}uential} Sequential $F$-test to obtain number of breaks when using hypothesis (3).

\hangpara
\texttt{csa(varlist)} specifies the variables with breaks which are added as cross-sectional averages. \texttt{xtbreak} calculates automatically the cross-sectional averages. 

\hangpara
\texttt{\underbar{csano}break()} is the same as \texttt{csa()} but for variables without a break.

\hangpara
\texttt{csd} implements \texttt{csa(\textit{w}) csanobreak(\textit{x})} automatically. For example, the variables in \textit{w} would enter with breaks, while those in \textit{x}, specified with the \texttt{\underbar{nobreakvar}iables(varlist1)}, enter without breaks.

\hangpara
\texttt{\underbar{kf}actors(varlist)} Variables in \varlist\ are known factors, variables in the data which are constant across the cross-sectional dimension. Examples are seasonal dummies or other observed common factors such as asset returns and oil prices. The factors in this list are affected by structural breaks in that their loadings change.

\hangpara
\texttt{\underline{nbkf}actors(varlist)} Same as above, but the factors in this list are not affected by structural breaks.

\hangpara
\texttt{skiph2} Skips hypothesis (2) ($H_0$: no break vs $H_1$: \(0 < s < s_{max}\) breaks) when running \texttt{xtbreak} without the \texttt{estimate} or \texttt{test} option.

\hangpara
\texttt{cvalue(level)} specifies the significance level to be used to estimate the number of breaks using the sequential test. For example \texttt{cvalue(0.99)} uses the 1\% significance level critical values to determine the number of breaks using the sequential test. See \texttt{level(\#)} for further details.

\hangpara
\texttt{strict} enforces strict behaviour of the sequential test to determine number of breaks.
Sequential test will stop once $F(s+1|s)$ is not rejected given a rejection of $F(s|s-1)$.

\hangpara
\texttt{\underline{max}breaks(\#)} limits maximum number of breaks when using the sequential test. 

\hangpara
\texttt{\underline{inv}erter(speed|precision|qr|chol|p|lu)} sets the inverter.  {\it speed} uses \texttt{mata invsym}, {\it precision}, {\it qr} (equivalent to precision; uses \texttt{mata qrinv}), 
{\it chol} uses \texttt{mata cholinv},
{\it p} uses \texttt{mata pinv}, or
{\it lu} uses \texttt{mata luinv}.
Choice of inverter has implications on speed and precision. 
For an overview see \mrefd{Solvers}.

\hangpara
\texttt{python} uses Python to calculate segment specific SSRs to improve speed. Requires Stata 16 or later, and Python packages, scipy, numpy, xarray and pandas. Numerical differences in the calculations may occur due to different matrix inverters and precision in Stata and Python. This option can only be used with balanced panels.

\hangpara
\texttt{noreweigh} can only be applied to unbalanced panels. As a default, \texttt{xtbreak} reweighs the time-unit specific errors used in the SSR by weights equal to total number of units in the sample divided by the number of unit observations in that period. Thus, it increases the SSR in segments with missing data. \texttt{noreweigh} avoids reweighting missing data.

\subsection{Stored values}

\noindent
\texttt{xtbreak estimate} stores the following in \texttt{e()}:

\begin{stresults}
	\stresultsgroup{Matrices} \\
	\stcmd{e(breaks)} &  Matrix with break dates
	& 
	\stcmd{e(CI)} & Confidence intervals with dimension 4 $\times$ number\_breaks \\
    \stcmd{e(SSRvec)} & Vector with SSRs for selection of break dates. Only available when \stcmd{breaks(1)} used. & 
    \stcmd{e(SSRvmat)} & Matrix with segment-specific SSR. Row indicates start, column indicates end of segment. 
\end{stresults}	

\begin{stresults}
	\stresultsgroup{Scalars} \\
	\stcmd{e(num\_breaks)} &  Number of breaks. If automatic detection used, then estimated number of breaks.
\end{stresults}	

\noindent
\texttt{xtbreak test} stores the following in \texttt{r()} for known breakpoints:

\begin{stresults}
	\stresultsgroup{Scalars} \\
	\stcmd{r(Wtau)} & Value of test statistic &  \qquad
	\stcmd{r(p)} & $p$-value from $F$ distribution
\end{stresults}	

\noindent For unknown breakpoints the following is stored:

\begin{stresults}
	\stresultsgroup{Scalars}\\
	\stcmd{r(supWtau)} & Value of the $\mathrm{sup}F(s)$ statistic (hypothesis (1)) & \qquad
	\stcmd{r(Dmax)} & Value of unweighted double maximum test statistic (hypothesis (2))\\
	\stcmd{r(WDmax)} & Value of weighted double maximum test statistic (hypothesis (2))&  \qquad
	\stcmd{r(f)} & Value of the $F(s+1|s)$ statistic (hypothesis (3))\\
	\stcmd{r(c90)} & Critival value at the 90\% level&  \qquad
	\stcmd{r(c95)} & Critival value at the 95\% level\\
	\stcmd{r(c99)} & Critival value at the 99\% level&
\end{stresults}

\subsection{Postestimation}

The following postestimation commands can be used after \texttt{xtbreak estimate}:

\begin{stsyntax}
	estat indicator [newvar]
\end{stsyntax}
\noindent creates an indicator variable \((1,...,1,2,...,2,...,s+1,...,s+1)'\) that specifies all break regimes. 

To split a varlist according to the estimated breakpoints:

\begin{stsyntax}
	estat split varlist
\end{stsyntax}

\noindent and to draw a scatter plot of the variable with break on the x-axis and the dependent variable in the y-axis:

\begin{stsyntax}
	estat scatter varname
\end{stsyntax}

\texttt{estat indicator} creates a new variable of the form \((1,...,1,2,...,2,...,s+1,...,s+1)'\) and where the value changes for each break regime. \texttt{estat split} splits the variables defined in varlist according to the breakdates.  \texttt{estat split} saves the names of the created variables in \textit{r(varlist)}.  \texttt{estat scatter} draws a scatter plot with the dependent variable on the y-axis and a variable with breaks defined in \textit{varname} on the x-axis.

\texttt{xtbreak estimate} stores information about segement-specific SSRs and the SSRs for different breakdates in \textbf{e()}. This information can be used to draw a time series plot of SSRs across potential breakdates with:

\begin{stsyntax}
    estat ssr , [\textit{tsline-options}]
\end{stsyntax}

\texttt{estat ssr} is only available after \texttt{xtbreak est y x , breaks(1)}. \textit{tsline-options} are any options permitted when using \texttt{tsline}, see \tsref{tsline} or \grefb{graph twoway tsline}.

\subsection{On the choice of factors in the panel case}

\texttt{xtbreak} is versatile in dealing with common factors, whether known (observed) or unknown (unobserved) through the options \texttt{csd}, \texttt{csa()}, \texttt{csanobreak()}, \texttt{kfactors()} and \texttt{nbkfactors()}. Unknown factors are estimated by the cross-sectional averages specified in \texttt{csa()} and \texttt{csanobreak()}, or alternatively in \texttt{csd}. Known factors can be dummy variables, or other observed variables which do not vary across units, and are defined by \texttt{kfactors()} and \texttt{nbkfactors()}. Whenever available, known factors should be included as they make it possible to estimate more unknown factors. These factors can be free of breaks or they can be allowed to break. 

Fixed effects are known factors but are not treated through the \texttt{kfactors()} and \texttt{nbkfactors()} commands, in order for \texttt{xtbreak}'s command structure to be similar to the rest of the Stata commands in the way that it deals with fixed effects. In particular, one has to specify whether \texttt{xtbreak} should estimate a model with or without overall constant/individual fixed effects, separately from the factor structure specification. In total, \texttt{xtbreak} supports five different models which are presented in Table 1. The default is a model with fixed effects that are not breaking.

\begin{table}[!h]
	\centering
\begin{tabular}{l l l}\hline\hline
Specification & Model & Options \\ \hline
Individual fixed effects & $\alpha_i+e_{i,t}$ & \texttt{noconstant}* \\
Fixed effects with breaks & $\alpha_{j,i}+e_{i,t}$ & \texttt{\underline{breakf}ixedeffects} \texttt{noconstant} \\
Overall constant & $\alpha+e_{i,t}$ & \texttt{\underline{nof}ixedeffects} \\
Constant with breaks & $\alpha_j+e_{i,t}$& \texttt{\underline{breakc}onstant \underline{nof}ixedeffects} \\
Nothing & $e_{i,t}$ & \texttt{\underline{noc}onstant \underline{nof}ixedeffects} \\ \hline\hline
\end{tabular}
\par
\begin{tabular}{l}
\small \emph{Notes}: * is the default. Option \texttt{noconstant} not necessary. $j=1,...s+1$. The remaining\\ \small observed and unobserved factors are in $e_{i,t}$ and require use of  \texttt{csd}, \texttt{csa()}, \texttt{csanobreak()},\\ \small \texttt{kfactors()} and \texttt{nbkfactors()} as appropriate.
\end{tabular}
\caption{Constant/fixed effects model specifications.}
\label{tab1}
\end{table}

The choice of deterministic model has implications for the analysis of structural breaks. In the overall constant model, the constant is treated as a regular regressor and we can test and detect breaks in it. In fact, the constant may be the only regressor. In the presence of fixed effects, \texttt{xtbreak} cannot be used without breaking regressors ($ w_{i,t} $), as it cannot detect or estimate breaks that affect only fixed effects and not the regressors. Once breaking regressors are included, then one has the choice to allow for breaking or non-breaking fixed effects.

\subsection{Unbalanced Panels}

\texttt{xtbreak} can be used with unbalanced panel data. Pure time series data ($N=1$) with gaps is not allowed. In the case of unbalanced panels, there is an appropriate adjustment in the degrees of freedom in the test statistics. It is assumed that missing data are missing completely at random and that all time periods have at least one observation.

In terms of break date estimation, \texttt{xtbreak} identifies breaks by minimising the SSR, see Equation \eqref{eq:breakdateestim}. If a value is missing, Stata removes the whole unit-time observation, and as such the number of summands in the SSR drop, which can lead to estimating break dates away from the true break. To avoid this behaviour, \texttt{xtbreak} reweighs the individual time specific residuals $\epsilon_{i,t}$ and the SSR is calculated as:
\begin{align}
SSR_{\tau_{S}} = \sum_{t \in \tau_{S}} \sum_{i \in N_t} \left(\frac{N}{N_t} \epsilon_{i,t}\right)^2
\end{align}
where $N$ are the number of total units and $N_t$ are the number of units for which data is non missing in time period $t$. $\frac{N}{N_t}$ scales the  residuals up when data is missing. The option \texttt{noreweigh} avoids the reweighing by setting $N_t = N$ independently from the number of non-missing observations.


\section{COVID--19 deaths and cases}

\subsection{Time series evidence for the whole US}

\subsubsection*{Main results}

In this section, we explain the use and options of \texttt{xtbreak}. We want to test if we can identify structural breaks in the relationship between the number of COVID--19 deaths and cases in the US in 2020 and 2021. This is an interesting topic because COVID--19's case fatality rate, which is the number of deaths from COVID--19 over the number of COVD--19 cases, was a key variable of medical and policy interest and consequently, the focus of many studies, see for example \cite{owidcoronavirus}.\footnote{Case fatality rate statistics can be found here: \href{https://ourworldindata.org/mortality-risk-covid}{https://ourworldindata.org/mortality-risk-covid}} The fatality rate can be time-varying due to, for example, country-wide changes in the capability of detecting the virus, lockdowns, population-wide vaccination programs, improvements in treatment and hospital capacity, and the emergence of new strains. All these events are massive in scale and can cause structural breaks. We use aggregate US weekly time series data on the number of deaths and new cases from the \href{https://data.cdc.gov/Case-Surveillance/United-States-COVID-19-Cases-and-Deaths-by-State-o/9mfq-cb36}{Centers for Disease Control and Prevention} (CDC). The dataset is also available on our \href{https://janditzen.github.io/xtbreak/}{GitHub} page.

We want to estimate the following model:
\begin{align}\label{eq:ea1}
	\Delta\mathrm{DEATHS}_t = \beta_0 + \delta_1 \Delta\mathrm{CASES}_{t-1}+ \delta_2 \Delta\mathrm{CASES}_{t-2}+ \delta_3 \Delta\mathrm{CASES}_{t-3} + e_t,
\end{align}
where \(\mathrm{DEATHS}_t\) and $\mathrm{CASES}_{t}$ are the reported deaths due to COVID--19 in week $t$, and the number of new cases for the entire US in the same week, respectively. We assume that on average a week lies between a positive test and a possible death.\footnote{Note that since we use weekly data, there can be up to 13 days between a positive test and a death. Monday is taken as the first day of a given week.} The data ranges from 27 January 2020 (beginning of week 4) to 29 August 2021 (end of week 34). $\Delta$ denotes first differencing, which is taken to ensure stationarity.\footnote{The regression \eqref{eq:ea1} can be seen as the static $\Delta\mathrm{DEATHS}_t = \beta_0 + \delta_1 \Delta\mathrm{CASES}_{t}+e_t$ model where we use lags of $\Delta\mathrm{CASES}_{t}$ to account for reverse causality concerns.} The model is a simplification, although it has been used elsewhere in the literature, for example in \cite{silverio2020} and \cite{fritz2022}.     

\begin{figure}
	\centering
	\includegraphics[width=\textwidth]{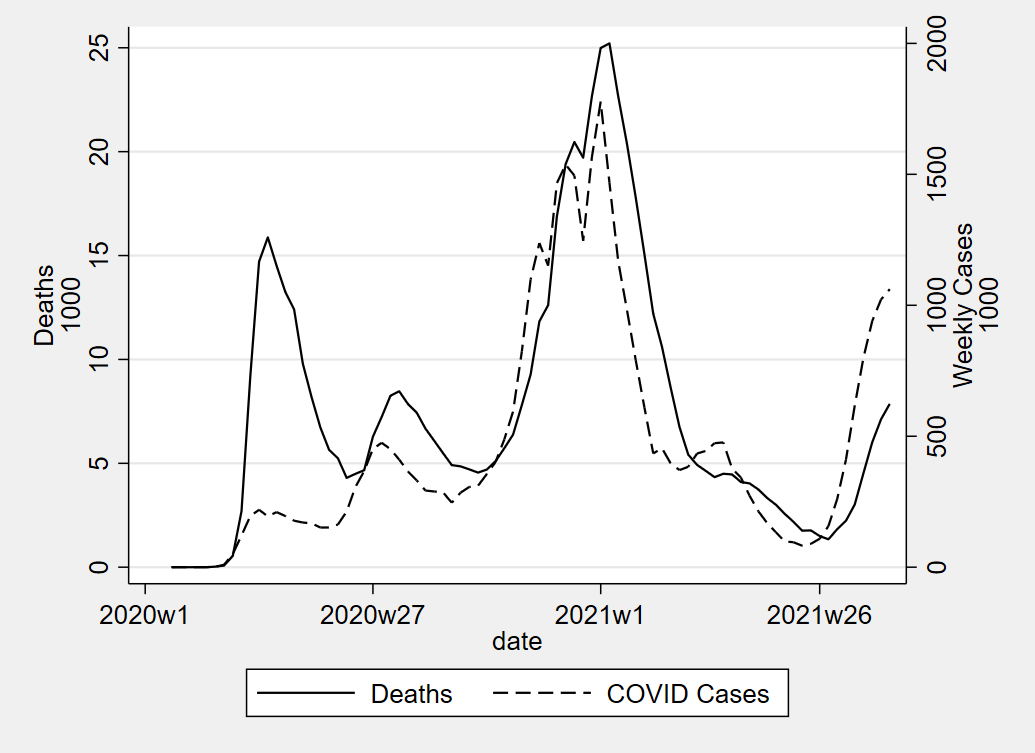}
	\caption{Plotting COVID--19 deaths and cases over time.}
	\label{fig:ExcessMortCovid}
\end{figure}

We want to test if the coefficients $\delta_1$, $\delta_2$ and $\delta_3$ are subject to structural breaks. There are several reasons for believing that the relationship between the number of new cases and deaths might have changed. At the beginning of the pandemic the understanding and best way to treat the disease was not developed. The numbers were also under-reported due to limits in testing capacity and reporting routines. With better testing capacity, reporting routines and treatments, the relationship is expected to change, especially after the first two waves, because of the knowledge gained. Starting in mid-December 2020, vaccines were introduced, and hence the relationship between the number of cases and deaths might be expected to have changed again. It therefore seems reasonable to expect at least two breaks.

Figure \ref{fig:ExcessMortCovid} plots the number of deaths and cases over time. The first thing to note is the striking difference between the number of deaths and cases in the first wave with the number of deaths being many times larger than the number of new cases. This difference is markedly smaller in the second wave but still the number of deaths is highest. The third wave was the worst in terms of numbers by far, but the number of deaths per new case was much lower than before. From about week 29 in 2020 the number of cases started to pick up again, and so did the number of deaths.

We can use \texttt{xtbreak} without any prior knowledge of the number of breaks or their exact dates. We use the following command line:

\begin{stlog}
	. xtbreak d.deaths d.L(1/3).cases
{\smallskip}
Test for multiple breaks at unknown breakdates
(Bai \& Perron. 1998. Econometrica)
H0: no break(s) vs. H1: 1 <= s <= 5 break(s)
{\smallskip}
                \HLI{17} Bai \& Perron Critical Values \HLI{17}
                     Test          1\% Critical     5\% Critical    10\% Critical
                  Statistic          Value            Value           Value
\HLI{80}
 UDmax               28.91             6.09            4.74            4.13
\HLI{80}
{\smallskip}
Sequential test for multiple breaks at unknown breakpoints
(Ditzen, Karavias \& Westerlund. 2024)
{\smallskip}
                \HLI{17} Bai \& Perron Critical Values \HLI{17}
                     Test          1\% Critical     5\% Critical    10\% Critical
                  Statistic          Value            Value           Value
\HLI{80}
 F(1|0)              28.51             6.09            4.66            4.03
 F(2|1)               5.47             6.59            5.24            4.64
 F(3|2)               2.78             6.92            5.61            4.99
 F(4|3)               2.70             7.33            5.87            5.23
 F(5|4)              19.84             7.49            6.05            5.45
\HLI{80}
Detected number of breaks: (min)          1               2               2
                           (max)          5               5               5
\HLI{80}
Null hypothesis rejected more than once after non-rejection.
 The detected number of breaks indicates the minimum and maximum
 number of breaks for which the null hypothesis is rejected.
{\smallskip}
Estimation of break points
                                            Number of obs       =     79
                                            SSR                 =     49.07
                                            Trimming            =      0.15
\HLI{80}
  \#      Index     Date                          [95\% Conf. Interval]
\HLI{80}
  1        15      2020w22                       2020w21        2020w23
  2        45      2020w52                       2020w51        2021w1 
\HLI{80}

\end{stlog}

\texttt{xtbreak} starts with hypothesis (2), that is, it tests the null hypothesis of no breaks against the alternative of an unknown number of breaks between 1 and $s_{max}$ breaks. This is the most powerful test and it does not need knowledge of the number of breaks. The maximum number of breaks is set to $s_{max} = 5$.\footnote{By default the maximum number of breaks is set to the maximum number of breaks permissible by the trimming parameter. The maximum permissible number of breaks depends on the minimal length of the subsamples considered, and hence on the trimming parameter $\epsilon$. The maximum number of breaks is given by the formula $\lceil 1/\epsilon\rceil-2 $, where $ \lceil \cdot \rceil $ is the smallest greater integer function. For $ \epsilon=0.15 $ the permissible number of breaks is $ 5 $, while for $ \epsilon=0.10 $ it is $8$ and for $ \epsilon=0.05 $ it is $ 18 $.}
The value of the UDmax test statistic is 28.91, well above the 1\% critical value and hence provides evidence for 1 up to 5 breaks. \texttt{xtbreak} then estimates the number of breaks by reporting the test value of Hypothesis (3) (H0: $s$ breaks vs $s+1$ breaks) at each step in the sequence and the appropriate critical values for the three basic significance levels. We see that 0 breaks is rejected in favour of 1 break and that 1 break is rejected in favour of 2 breaks, at the 5\% significance level, but then when testing 2 breaks against 3 or more breaks the test is no longer able to reject.\footnote{\texttt{xtbreak} does not stop and reports results for all tests, up to the maximum numbers of breaks. This aims to provide a broader view of the sample properties as individual tests can suffer from Type 1 errors. In this example we see that the null hypothesis of 4 breaks against the alternative of 5 breaks is strongly rejected. This can be investigated in a robustness analysis, but we do not pursue it here. In such cases, one should keep an eye out on the magnitude of breaking coefficients as the method may be detecting small breaks which are not economically significant, see the empirical application in \cite{DKW2021}. An alternative explanation could be that there are multiple breaks with opposing magnitudes that ``hide'' each other, until the number of breaks is set high enough to disentangle the opposing signs - and then the breaks become visible. This is why it is important to test hypothesis (2) before applying the sequential tests. However, this last scenario is unlikely in the current example because it is expected that the fatality rate only drops in time.}\footnote{The detected minimum and maximum number of breaks correspond to the first and last $F(s+1|s)$ test rejections; at the 1\% confidence level the minimum detected number of breaks is $1$ because $F(1|0)$ rejects and the maximum is $5$ because $F(5|4)$ rejects. It is possible to stop this process at the first non-rejection at the specified level using the option \texttt{strict}.} We therefore conclude that there are 2 breaks. \texttt{xtbreak} then proceeds to report the estimated break dates and the associated confidence intervals. The first break is estimated to week 22 of 2020 (fourth week of May), while the second is estimated at the end of year 2020. The confidence intervals on the both breaks span only a couple of weeks, suggesting that the estimates are precise. The confidence interval width is a function of the magnitude of the break and short confidence intervals hint at larger breaks.

\begin{figure}
	\centering
	\includegraphics[width=\textwidth]{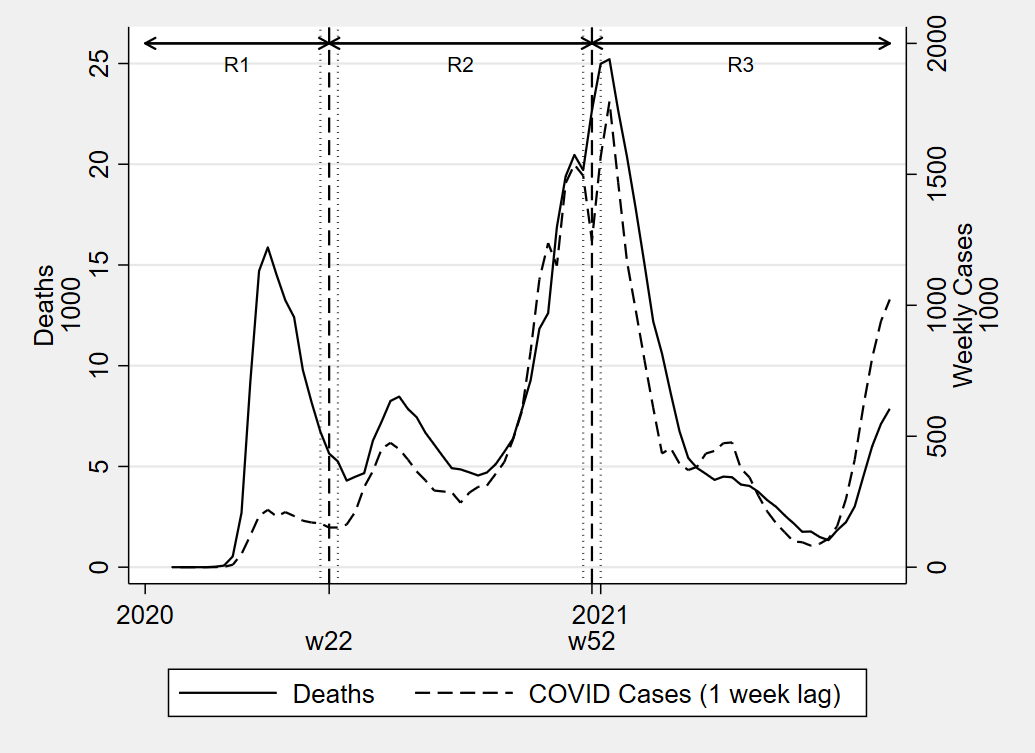}
	\caption{Plotting estimated breaks (dashed lines), 95\% confidence intervals (dotted lines), deaths and lagged cases.}
	\label{fig:DeathsCI}
\end{figure}

\texttt{xtbreak} saves the values of the breakpoints in \texttt{e(breaks)} and the confidence intervals in \texttt{e(CI)}. The matrices contain the index number $t\in \{1,...,T\}$ of the both breaks and confidence interval bounds. We use this information to draw Figure \ref{fig:DeathsCI}, in which the estimated break dates and their 95\% confidence intervals are plotted in the same graph as the deaths and the lagged number of cases.\footnote{The code for the plot is available in the accompanying do file.}

Next we can use the \texttt{estat} function of \texttt{xtbreak} to generate new variables for each break regime and run an OLS regression:

\begin{stlog}
	. estat split
New variables created: LD_cases1 LD_cases2 LD_cases3 
>   L2D_cases1 L2D_cases2 L2D_cases3 
>   L3D_cases1 L3D_cases2 L3D_cases3
{\smallskip}
. reg d.deaths `r(varlist)'
{\smallskip}
      Source {\VBAR}       SS           df       MS      Number of obs   =        79
\HLI{13}{\PLUS}\HLI{34}   F(9, 69)        =     25.19
       Model {\VBAR}   161.21305         9  17.9125611   Prob > F        =    0.0000
    Residual {\VBAR}  49.0738642        69  .711215423   R-squared       =    0.7666
\HLI{13}{\PLUS}\HLI{34}   Adj R-squared   =    0.7362
       Total {\VBAR}  210.286914        78  2.69598608   Root MSE        =    .84334
{\smallskip}
\HLI{13}{\TOPT}\HLI{64}
    D.deaths {\VBAR} Coefficient  Std. err.      t    P>|t|     [95\% conf. interval]
\HLI{13}{\PLUS}\HLI{64}
   LD_cases1 {\VBAR}   .0625909    .009525     6.57   0.000      .043589    .0815928
   LD_cases2 {\VBAR}   .0014608   .0015401     0.95   0.346    -.0016116    .0045332
   LD_cases3 {\VBAR}   .0080629   .0012074     6.68   0.000     .0056542    .0104717
  L2D_cases1 {\VBAR}   .0298979   .0121457     2.46   0.016      .005668    .0541279
  L2D_cases2 {\VBAR}   .0037864   .0018352     2.06   0.043     .0001251    .0074476
  L2D_cases3 {\VBAR}  -.0005085   .0012946    -0.39   0.696    -.0030911    .0020741
  L3D_cases1 {\VBAR}  -.0171167   .0096604    -1.77   0.081    -.0363887    .0021553
  L3D_cases2 {\VBAR}   .0065269    .001718     3.80   0.000     .0030995    .0099543
  L3D_cases3 {\VBAR}   .0030002   .0011672     2.57   0.012     .0006717    .0053288
       _cons {\VBAR}  -.1963545   .1030999    -1.90   0.061    -.4020332    .0093243
\HLI{13}{\BOTT}\HLI{64}
{\smallskip}
. nlcom   (Regime1: _b[LD_cases1] +  _b[L2D_cases1] + _b[L3D_cases1]) ///
> (Regime2: _b[LD_cases2] +  _b[L2D_cases2] + _b[L3D_cases2]) ///
> (Regime3: _b[LD_cases3] +  _b[L2D_cases3] + _b[L3D_cases3])
{\smallskip}
     Regime1: _b[LD_cases1] +  _b[L2D_cases1] + _b[L3D_cases1]
     Regime2: _b[LD_cases2] +  _b[L2D_cases2] + _b[L3D_cases2]
     Regime3: _b[LD_cases3] +  _b[L2D_cases3] + _b[L3D_cases3]
{\smallskip}
\HLI{13}{\TOPT}\HLI{64}
    D.deaths {\VBAR} Coefficient  Std. err.      z    P>|z|     [95\% conf. interval]
\HLI{13}{\PLUS}\HLI{64}
     Regime1 {\VBAR}   .0753721   .0089054     8.46   0.000     .0579178    .0928265
     Regime2 {\VBAR}    .011774   .0020155     5.84   0.000     .0078238    .0157243
     Regime3 {\VBAR}   .0105547   .0014136     7.47   0.000      .007784    .0133254
\HLI{13}{\BOTT}\HLI{64}

\end{stlog}

The long-run multipliers within each regime are given by $\delta_{LR,j}:=\delta_{1,j}+\delta_{2,j}+\delta_{3,j}$ for $j=1,2,3$. They are reported using the \texttt{nlcom} command which also provides  confidence intervals. The $\delta_{LR,j}$ capture the total effect in time that an additional 1,000 COVID--19 cases have on deaths, in regime $j$. The estimate for the first regime $\hat \delta_{LR,1}$ covering the period from week 4 of 2020 to week 22 of 2020 suggests that for each additional 1,000 cases of COVID--19, on average 75 people died. According to Figure \ref{fig:DeathsCI}, the end date of this regime coincides with the end of the first wave. This estimate is quite high, which can be explained by the relatively small number of tests conducted at that time. The estimated effect of the number of cases is much lower in the second regime, which stretches from week 23 of 2020 to week 51 of 2020. Only 12 out of an additional 1,000 infected died. The coefficient of the third regime, lasting from week 1 in 2021 to the end of the sample, is almost the same, where 11 out of an additional 1,000 infected died. This is probably in part due to the vaccination rollout, which began in week 51 of 2020.

\subsubsection*{Additional results and discussions}

This section is divided in two; (i) additional test results and (ii) additional estimation results. The aim is to demonstrate how the command can be used for specific tests and estimations, as opposed to the automatic way used above. We begin by considering additional test results, starting with a test of hypothesis (1). Specifically, we test the null hypothesis of no breaks against the alternative of a break in week 22 in 2020 and another break in week 52 in 2020 using the regular Chow $F$-test. There are the results:

\begin{stlog}
	. xtbreak test d.deaths d.L(1/3).cases , hypothesis(1) breakpoints(2020W22 2020w52, fmt(tw))
Test for multiple breaks at known breakdates
(Bai \& Perron. 1998. Econometrica)
H0: no breaks vs. H1: 2 break(s)
{\smallskip}
 F             =       19.95
 p-value (F)   =        0.00
{\smallskip}

\end{stlog}

The test value is 19.95, which is way out in the critical region of the $F-$distribution. We can therefore comfortably reject the null of no breaks. To test the same hypothesis but with unknown break dates, we specify \texttt{breaks(2)}. The results look as follows:

\begin{stlog}
	. xtbreak test d.deaths d.L(1/3).cases , hypothesis(1) breaks(2)
{\smallskip}
Test for multiple breaks at unknown breakdates
(Bai \& Perron. 1998. Econometrica)
H0: no break(s) vs. H1: 2 break(s)
{\smallskip}
                \HLI{17} Bai \& Perron Critical Values \HLI{17}
                     Test          1\% Critical     5\% Critical    10\% Critical
                  Statistic          Value            Value           Value
\HLI{80}
 supF                19.95             4.82            4.00            3.58
\HLI{80}
Estimated break points: 2020w22 2020w52
Trimming: 0.15

\end{stlog}

The test value is identical to before, as is the conclusion to reject the null hypothesis, which is to be expected because the given break dates were set to the estimated breaks obtained earlier. The difference is that when the breaks are treated as unknown, the critical values come from a non-standard distribution, because the statistic is the supremum of $F$-tests over all possible break dates, as these are determined by the trimming parameter. These critical values are more ``honest'' than those used for the Chow test, as they account for the fact that the breaks are unknown. As a part of the test results, \texttt{xtbreak} reports the estimated break dates used to construct the test, and we can see that they coincide with those obtained earlier.

We now test the null of no breaks against the alternative of up to 5 breaks. This is an example of a test of hypothesis (2), the results of which are presented here below. As expected, the null hypothesis is firmly rejected.

\begin{stlog}
	. xtbreak test d.deaths d.L(1/3).cases , hypothesis(2) breaks(5) 
{\smallskip}
Test for multiple breaks at unknown breakdates
(Bai \& Perron. 1998. Econometrica)
H0: no break(s) vs. H1: 1 <= s <= 5 break(s)
{\smallskip}
                \HLI{17} Bai \& Perron Critical Values \HLI{17}
                     Test          1\% Critical     5\% Critical    10\% Critical
                  Statistic          Value            Value           Value
\HLI{80}
 UDmax               28.91             6.09            4.74            4.13
\HLI{80}
Trimming: 0.15

\end{stlog}

Next, we test the null of 3 breaks against the alternative of 4 breaks, which is an example of hypothesis (3). We use the options \texttt{hypothesis(3)} and \texttt{breaks(4)} to specify that there are 4 breaks under the alternative. The results presented below suggest that we are unable to reject the null, which is consistent with the estimated number of breaks.

\begin{stlog}
	. xtbreak test d.deaths d.L(1/3).cases , hypothesis(3) breaks(4)
{\smallskip}
Test for multiple breaks at unknown breakpoints
(Bai \& Perron. 1998. Econometrica)
H0: 3 vs. H1: 4 break(s)
{\smallskip}
                \HLI{17} Bai \& Perron Critical Values \HLI{17}
                     Test          1\% Critical     5\% Critical    10\% Critical
                  Statistic          Value            Value           Value
\HLI{80}
 F(s+1|s)*            2.70             7.33            5.87            5.23
\HLI{80}
* s = 3
Trimming: 0.15

\end{stlog}

 The option \texttt{sequential} repeats the hypothesis (3) test sequentially starting from no breaks under the null, up to the specified number in \texttt{breaks}. Setting \texttt{breaks(5)} returns the sequential test results reported earlier.

\begin{stlog}
	xtbreak test d.deaths d.L(1/3).cases , hypothesis(3) breaks(5) sequential
\end{stlog}

As a final test exercise, we consider two changes to the model. We begin by investigating how the above results are affected if we allow for breaks in the constant. To do so we use the sequential test but add the option \texttt{breakconstant};

\begin{stlog}
	xtbreak test d.deaths d.L(1/3).cases, breakconstant
\end{stlog}

To save space we omit the output, but we briefly describe it;  \texttt{xtbreak} finds five breaks. Note also that the options \texttt{hypothesis(3)} and \texttt{sequential} are the defaults, so we have omitted from the command. We further investigate if the break is only in the constant, and not in the number of cases. We keep the option \texttt{breakconstant} and move the variable \texttt{L.cases} to the option \texttt{nobreakvar(L.cases)};

\begin{stlog}
	xtbreak test d.deaths , breakconstant nobreakvar(d.L(1/3).cases)
\end{stlog}

Now \texttt{xtbreaks} detects no breaks at the 5\% significance level, meaning that the breaks are driven from changes in the slope coefficients. 

We end this section with some comments on the estimation results. The break date results reported earlier can be obtained using the option \texttt{xtbreak estimate}. The appropriate command line is the following:

\begin{stlog}
	xtbreak estimate d.deaths d.L(1/3).cases , breaks(2)
\end{stlog}

As an illustration of the estimated regression model, we can draw a scatter plot with different symbols for the observations within each regime. The plot can be created using \texttt{xtbreak estat}:

\begin{stlog}
	. estat scatter d.L.cases , ytitle("Change in Deaths in 1000s") xtitle("Change in Cases in 1000s") ///
>         autolegend(pos(6) cols(3)) scheme(sj) name(xtbreak_estat, replace)
{\smallskip}

\end{stlog}

\begin{figure}
	\centering
	\includegraphics[width=\textwidth]{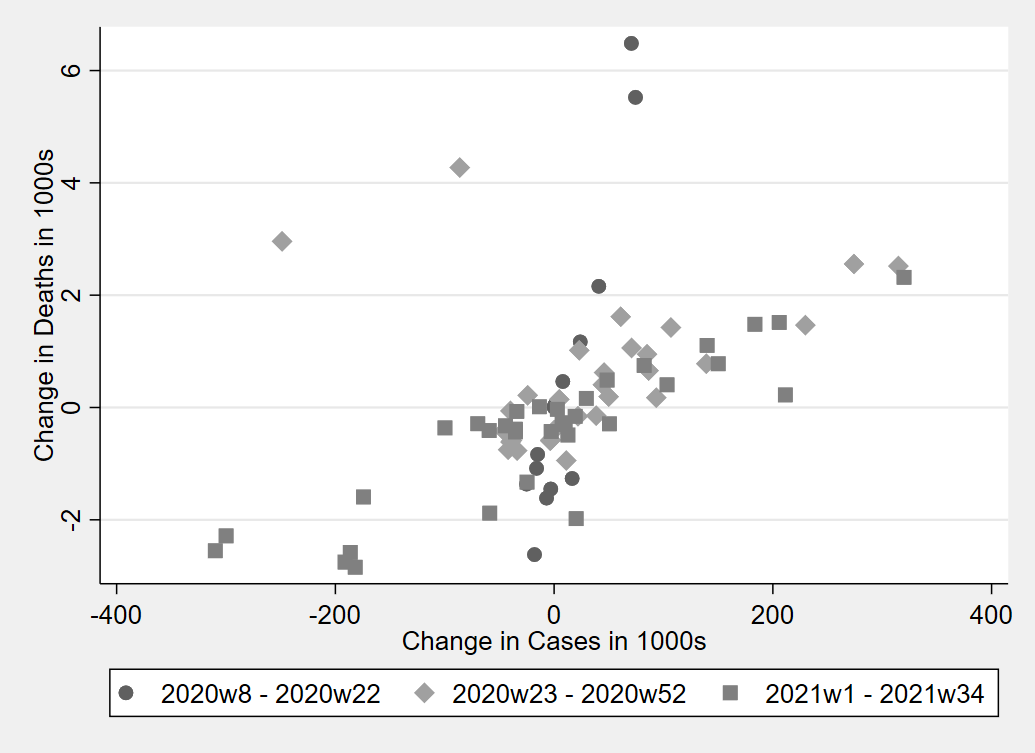}
	\caption{Scatter plotting lagged cases against deaths by regime.}
	\label{fig:estatScatter}
\end{figure}

The plot is displayed in Figure \ref{fig:estatScatter}. The different markers represent the observations within each regime, and the shape of the observations is indicative of the strength of the estimated linear relationships and of their slopes. The first regime is described by dots, which appear in an almost vertical line. The second and third regimes are described by rhombi and squares respectively, which mostly lie along the 45-degree line. In particular, we see that the slope in the first regime differs quite markedly when compared to the other two, which confirms the findings from the OLS regression. \grefb{graph} options can be passed through. Option \texttt{autolegend(\textit{legend\_options})} automatically creates the legend labels for each segment. \textit{legend\_options} are further options passed to \texttt{legend}, as done here to control the placement and number of columns of the legend.

\subsection{State level panel evidence}

\subsubsection*{Main results}

In this section, we use the same US data as in the previous section; however, instead of aggregating the data up to the country level, we use data for all 50 US states, the District of Columbia, New York City, overseas territories and three countries in free association with the US.\footnote{The overseas territories and three countries are; American Samoa, Guam, the Commonwealth of the Northern Mariana Islands, Puerto Rico, the US Virgin Islands, the Federated States of Micronesia, Republic of the Marshall Islands, and Republic of Palau.}

Similarly to before, report some results based on the sequential test to estimate the number of breaks and the break dates. We begin with the automatic \texttt{xtbreak} command that tests hypothesis (3) sequentially. The default panel data case assumes no serial correlation or cross-section dependence, has no constant but includes fixed effects. The distributed lag model employed here captures serial corrlation and thus we employ only  heteroskedasticity robust standard errors. Additionally, we employ a smaller trimming of 10\%.\footnote{Thicker trimming offers better small sample properties for the test statistics. The availability of rich datasets allows the use of thinner trimming, which permits more breaks and also allows breaks to be closer to each other. More breaks do not increase the computational burder per se, due to the dynamic programming algorithms employed. However, more breaks can mean more sequential tests of hypotheses $F(s+1|s)$ which will ultimately delay the automatic version of \texttt{xtbreak}. Limits on the number of h(3) tests can be introduced by i) using \texttt{xtbreak test} directly, and ii) by adding options such as \texttt{strict} and \texttt{maxbreaks}.} \texttt{xtbreak} automatically detects if a panel data model is used, and thus the syntax remains the same. These are the results:

\begin{stlog}
	. xtbreak d.deaths d.L(1/3).cases, vce(hc) trim(0.1) 
{\smallskip}
Test for multiple breaks at unknown breakdates
(Ditzen, Karavias \& Westerlund. 2024)
H0: no break(s) vs. H1: 1 <= s <= 9 break(s)
{\smallskip}
                \HLI{17} Bai \& Perron Critical Values \HLI{17}
                     Test          1\% Critical     5\% Critical    10\% Critical
                  Statistic          Value            Value           Value
\HLI{80}
 UDmax               13.60             6.25            4.95            4.42
\HLI{80}
{\smallskip}
Sequential test for multiple breaks at unknown breakpoints
(Ditzen, Karavias \& Westerlund. 2024)
{\smallskip}
                \HLI{17} Bai \& Perron Critical Values \HLI{17}
                     Test          1\% Critical     5\% Critical    10\% Critical
                  Statistic          Value            Value           Value
\HLI{80}
 F(1|0)              11.90             6.24            4.87            4.26
 F(2|1)               9.31             6.78            5.51            4.85
 F(3|2)               8.50             7.20            5.81            5.21
 F(4|3)              10.76             7.45            5.99            5.49
 F(5|4)               1.14             7.65            6.20            5.65
 F(6|5)               4.82             7.79            6.34            5.78
 F(7|6)               1.24             7.84            6.42            5.89
 F(8|7)               2.20             7.90            6.54            5.98
 F(9|8)               2.44             7.93            6.65            6.12
\HLI{80}
Detected number of breaks:                4               4               4
\HLI{80}
The detected number of breaks indicates the highest number of
 breaks for which the null hypothesis is rejected.
{\smallskip}
Estimation of break points
                                            Number of obs       =   4740
                                            Number of Groups    =     60
                                            Obs per group       =     79
                                            SSR                 =     13.88
                                            Trimming            =      0.10
\HLI{80}
  \#      Index     Date                          [95\% Conf. Interval]
\HLI{80}
  1        7       2020w14                       2020w13        2020w15
  2        14      2020w21                       2020w20        2020w22
  3        46      2021w1                        2020w52        2021w2 
  4        53      2021w8                        2021w7         2021w9 
\HLI{80}

\end{stlog}

At a 5\% significance level we find four breaks which are estimated in weeks 14, 21 and 46 of 2020, and in week 8 of 2021. The second and third breaks are remarkably close to the two breaks found in the single-time series analysis above.\footnote{The model then can be estimated by using the commands \texttt{estat split} to generate the breaking variables and \texttt{xtreg d.deaths `r(varlist)', fe} to run a fixed effects regression.} We will comment on the break dates below. Given that COVID--19 waves impact multiple states at the same time there may be dependence across states. We address this issue by augmenting the model with the cross-section average of the lagged number of differenced cases.\footnote{The CCE methodology requires that the cross-section average of each regressor is used, which in this case translates to the cross-section averages of the three lags. This can be done by options \texttt{csd} or explicitly by \texttt{d.l(1/3).cases}. However, we point out that the model employed is a distributed lag model with high persistence in the regressors and thus the three cross section averages introduced by \texttt{d.l(1/3).cases} are highly correlated, containing almost the same information $\bar x_{t-1} \approx \bar x_{t-2} \approx \bar x_{t-3}$. In this case it makes sense to use only one cross-sectional average, which we implement with the option \texttt{csa(d.l.cases)}.} Additionally, we employ the option \texttt{strict} which uses the sequential test until it does not reject the null at the required significance level. The option \texttt{strict} provides the consistent number of breaks estimator of Theorem 3.2 in \cite{DKW2021}. Another option used is \texttt{skiph2} which skips the test of no breaks against \(1 < s < s_{max}\) breaks to save space, as it strongly rejects. We employ the heteroskedasticity and autocorrelation robust variance estimator of \cite{wpn19} due to its excellent small sample properties. The results look as follows:

\begin{stlog}
	. xtbreak d.deaths d.L(1/3).cases,  csa(d.l.cases) vce(wpn) trim(0.1) skiph2
{\smallskip}
Sequential test for multiple breaks at unknown breakpoints
(Ditzen, Karavias \& Westerlund. 2024)
{\smallskip}
                \HLI{17} Bai \& Perron Critical Values \HLI{17}
                     Test          1\% Critical     5\% Critical    10\% Critical
                  Statistic          Value            Value           Value
\HLI{80}
 F(1|0)               6.32             6.24            4.87            4.26
 F(2|1)              68.68             6.78            5.51            4.85
 F(3|2)              73.44             7.20            5.81            5.21
 F(4|3)               9.10             7.45            5.99            5.49
 F(5|4)               4.50             7.65            6.20            5.65
 F(6|5)              13.82             7.79            6.34            5.78
 F(7|6)              10.33             7.84            6.42            5.89
 F(8|7)              10.40             7.90            6.54            5.98
 F(9|8)               8.04             7.93            6.65            6.12
\HLI{80}
Detected number of breaks: (min)          4               4               4
                           (max)          9               9               9
\HLI{80}
Null hypothesis rejected more than once after non-rejection.
 The detected number of breaks indicates the minimum and maximum
 number of breaks for which the null hypothesis is rejected.
{\smallskip}
Estimation of break points
                                            Number of obs       =   4740
                                            Number of Groups    =     60
                                            Obs per group       =     79
                                            SSR                 =     10.89
                                            Trimming            =      0.10
\HLI{80}
  \#      Index     Date                          [95\% Conf. Interval]
\HLI{80}
  1        7       2020w14                       2020w13        2020w15
  2        14      2020w21                       2020w20        2020w22
  3        45      2020w52                       2020w51        2021w1 
  4        53      2021w8                        2021w7         2021w9 
\HLI{80}
Cross-section averages:
  with breaks: LD.cases

\end{stlog}

The option \texttt{strict} has a default significance level of 5\%. For this significance level the $F(5|4)$ test does not reject and the sequential procedures stops yielding an estimate of 4 breaks. Notably the breaks are the same as before. 

\subsubsection*{Additional results and discussions}

The estimated breaks take place in week 14 of 2020, in week 21 of 2020, in week 52 of 2020, and in week 8 of 2021. The confidence intervals for all five breaks are narrow. When we compare these results to those reported earlier for the time series data set for the whole US, we see that the second and third breaks coincide. Hence, for two of the break dates, the panel data evidence reinforces the time series evidence reported earlier. However, the panel data results also suggest that two breaks are not enough and that there is a need to account for an early break in the week 14 of 2020 and for a fourth break in week 8 of 2021. The fact that the panel toolbox detects two breaks which are not picked up by the time series analysis could be due to the gain in accuracy obtained by using the larger panel data set (see \citealp{DKW2021}).

The command \texttt{estat split} generates the breaking regressors which will be used as independent variables. The model is estimated by CCE using \texttt{xtdcce2} \citep{Ditzen2018xtdcce2,Ditzen2021}, as the CCE estimator allows for interactive effects which capture cross-section dependence. Same as above, we report the long-run multipliers for each regime. 
The estimated multiplier in the first regime, which corresponds to the period of the first wave, is the highest. This is because there was little medical knowledge about the virus at that time, limited preparedness, and limited capacity in detecting cases. The latter can be seen in Figure 4. The multipliers drop in the second regime as the first wave dissipates and more testing becomes available. The third regime includes the second and partly the third wave appearing in October 2020. The multiplier further drops here. The fourth regime includes the peak of the third wave. Here cases and deaths are close to each other due to extensive tracing programs employed across states and also mass vaccinations. The multiplier remains almost the same in the fifth and final regime. This is an example of a break which is statistically significant but may not be economically significant. For this reason we do not consider more breaks in the analysis.\footnote{In large datasets an overspecified model with more breaks is preferable to an underspecified model, because in the latter the estimators are inconsistent (see \citealp{DKW2021}).}


\begin{stlog}
. qui xtdcce2 d.deaths `r(varlist)', pooled(`r(varlist)') cr(d.L.cases) pooledvce(wpn)
{\smallskip}
. nlcom   (Regime1: _b[LD_cases1] +  _b[L2D_cases1] + _b[L3D_cases1]) ///
> (Regime2: _b[LD_cases2] +  _b[L2D_cases2] + _b[L3D_cases2]) ///
> (Regime3: _b[LD_cases3] +  _b[L2D_cases3] + _b[L3D_cases3]) ///
> (Regime4: _b[LD_cases4] +  _b[L2D_cases4] + _b[L3D_cases4]) ///
> (Regime5: _b[LD_cases5] +  _b[L2D_cases5] + _b[L3D_cases5]) ,  post
{\smallskip}
     Regime1: _b[LD_cases1] +  _b[L2D_cases1] + _b[L3D_cases1]
     Regime2: _b[LD_cases2] +  _b[L2D_cases2] + _b[L3D_cases2]
     Regime3: _b[LD_cases3] +  _b[L2D_cases3] + _b[L3D_cases3]
     Regime4: _b[LD_cases4] +  _b[L2D_cases4] + _b[L3D_cases4]
     Regime5: _b[LD_cases5] +  _b[L2D_cases5] + _b[L3D_cases5]
{\smallskip}
\HLI{13}{\TOPT}\HLI{64}
    D.deaths {\VBAR} Coefficient  Std. err.      z    P>|z|     [95\% conf. interval]
\HLI{13}{\PLUS}\HLI{64}
     Regime1 {\VBAR}   .1331224   .0244026     5.46   0.000     .0852942    .1809506
     Regime2 {\VBAR}   .0555998   .0163378     3.40   0.001     .0235783    .0876213
     Regime3 {\VBAR}   .0150175   .0011888    12.63   0.000     .0126876    .0173475
     Regime4 {\VBAR}   .0104796   .0015258     6.87   0.000      .007489    .0134701
     Regime5 {\VBAR}   .0107557   .0010314    10.43   0.000     .0087343    .0127772
\HLI{13}{\BOTT}\HLI{64}

\end{stlog}


\begin{figure}
	\centering
	\includegraphics[width=\textwidth]{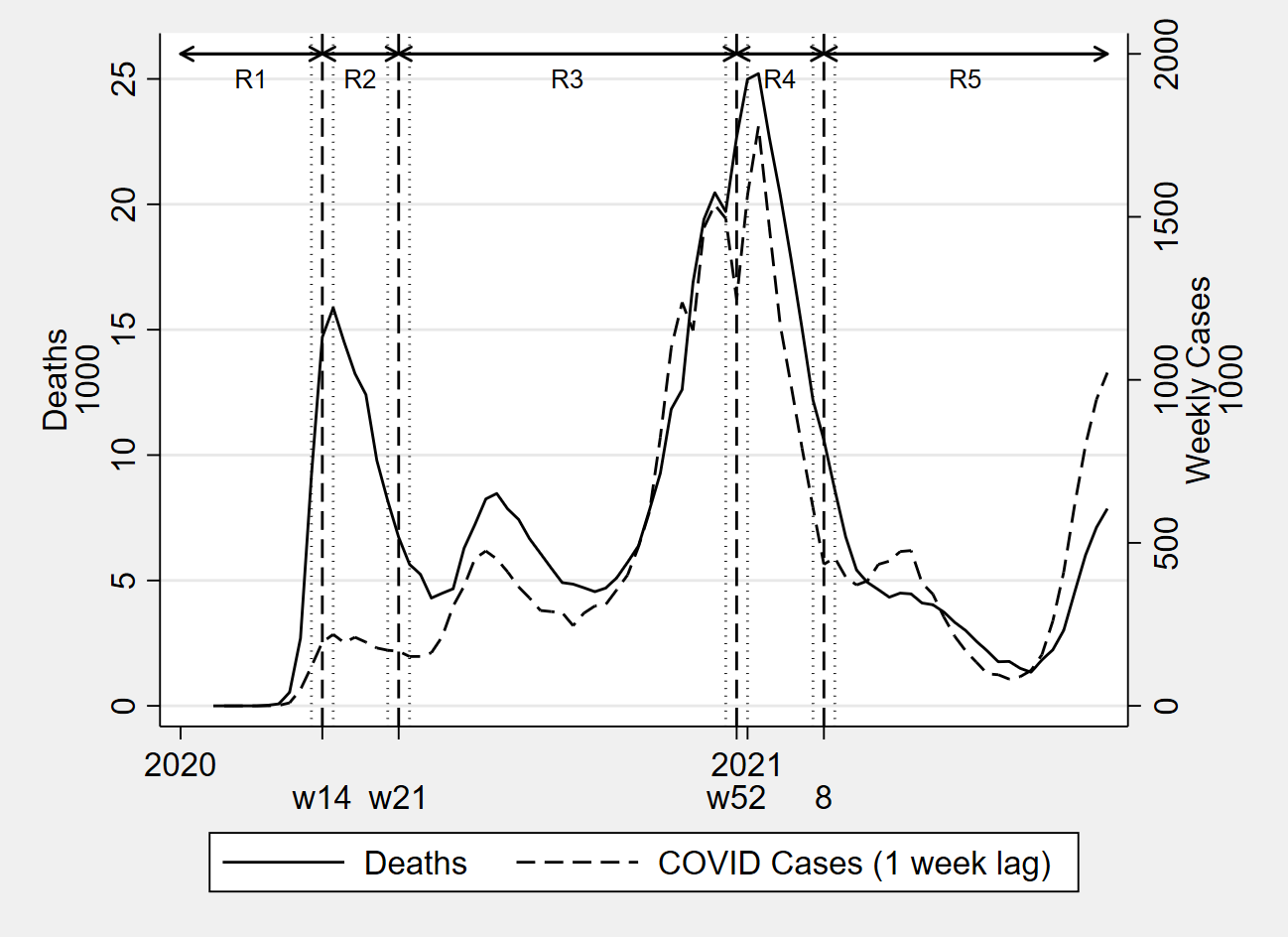}
	\caption{Plotting estimated breaks (dashed lines), 95\% confidence intervals (dotted lines), deaths and lagged cases. The arrows on top indicate the Regimes. }
	\label{fig:DeathsCIPanel2}
\end{figure}






\section{Consumer Confidence and Leader Approval Rating}

Another relationship that can potentially suffer from structural breaks is the one between consumer confidence and the approval rating of a country's leader. In the US, there is now a large literature analysing the determinants of presidential ratings (see for example \cite{BERLEMANN} for an extensive review). The determinant of interest here is consumer confidence, a statistical measure of consumer feelings on the state of the economy and their financial situation. Confidence in the economy should translate into high presidential ratings, although the relationship itself may be unstable at times with structural breaks occurring for various reasons, including political instability and other significant events, an example being the inauguration of President Barack Obama (\cite{small}). Another dimension of interest when studying such relationships is their cross-country heterogeneity, which is the outcome of unique historical processes shaping institutions and culture. In the words of \cite{Nannestad}, ``approval relationships have shown a disappointing lack of stability both over time and across countries''.

The present study is motivated by the above discussion and will attempt to discover breaks in a set of eight countries in which leader approval ratings were observed monthly from January 1990 to December 2021. The data on approval ratings were obtained from the EAP 3.0 Database \citep{EAP3}, while the data on consumer confidence were taken from the OECD.\footnote{In the panel case the validity of critical values requires theoretically that $N>>T$. However, Monte Carlo evidence in \cite{Ditzen2021} show that this requirement is not critical. See also the discussion at the end of Section 3.1.} The two variables are denoted as \textit{approval} and \textit{CCI}. As the time dimension is long, we set the trimming parameter to 5\%. Additionally, we control for breaks due to elections which can be seen as known break dates. Elections dates for the years post 2000 come from the \cite{EG2023} and are hand-filled for the years before 2000. To account for political environment right before and after an election, the dummy variable \textit{ElectionQ} equals one in the month prior and for the two months after an election. We analyse first the whole panel and later on each country separately.

In the estimations below we allow unobserved heterogeneity to account for country-specific variation and use first differences of the variables to ensure stationarity. As the time dimension is large, a large number of possible SSRs has to be estimated which takes a considerable amount of time.\footnote{The trimming of 5\% and the 383 time periods imply a minimal segment length of $h = 20$ periods. The number of possible segments is $T(T+1)/2 - (h-1)T + (h-2)(h-1)/2 -h^2m(m+1)/2$ with m the number of breaks \citep{Bai2003}. For $m=1$ this amounts to over 65,000 estimations.} In order to save time, we restrict the number of breaks under the null hypothesis for hypothesis (2) to $s_{max}=5$ using the options \texttt{maxbreaks(5)} and \texttt{strict} for the sequential test and invoke the option \texttt{python}.\footnote{We did not find more than 5 breaks for the entire panel or any individual country. Using the option \texttt{python} reduces the computation time significantly.} With the last option, the sequential test will stop once the null hypothesis $F(s+1|s)$ is not rejected given a rejection of $F(s|s-1)$ at the default 5\% significance level. 

\begin{stlog}
. xtbreak  d.approval d.CCI , trim(0.05) nobreakvar(ElectionQ)  strict maxbreaks(5) python
{\smallskip}
Test for multiple breaks at unknown breakdates
(Ditzen, Karavias \& Westerlund. 2024)
H0: no break(s) vs. H1: 1 <= s <= 5 break(s)
{\smallskip}
                \HLI{17} Bai \& Perron Critical Values \HLI{17}
                     Test          1\% Critical     5\% Critical    10\% Critical
                  Statistic          Value            Value           Value
\HLI{80}
 UDmax              13.91            13.74           10.17            8.78
\HLI{80}
{\smallskip}
Sequential test for multiple breaks at unknown breakpoints
(Ditzen, Karavias \& Westerlund. 2024)
{\smallskip}
                \HLI{17} Bai \& Perron Critical Values \HLI{17}
                     Test          1\% Critical     5\% Critical    10\% Critical
                  Statistic          Value            Value           Value
\HLI{80}
 F(1|0)              12.63            13.58            9.63            8.02
 F(2|1)              26.99            15.03           11.14            9.56
 F(3|2)               5.32            15.62           12.16           10.45
\HLI{80}
Detected number of breaks:                2               2               2
\HLI{80}
The detected number of breaks indicates the highest number of
 breaks for which the null hypothesis is rejected.
{\smallskip}
Estimation of break points
                                            Number of obs       =   3064
                                            Number of Groups    =      8
                                            Obs per group       =    383
                                            SSR                 =  34584.14
                                            Trimming            =      0.05
\HLI{80}
  \#      Index     Date                          [95\% Conf. Interval]
\HLI{80}
  1        344     2018m9                        2017m3         2020m3 
  2        363     2020m4                        2019m2         2021m6 
\HLI{80}

\end{stlog}

The UDmax test is rejected at the 10\% level, suggesting at least one break in the relationship. Such evidence is rather weak given the size of the sample, however, looking at the sequential statistic results, there is strong evidence in favour of two breaks, at all significance levels. The two breaks are found in September 2018 and April 2020. The confidence intervals of the breaks span in both cases more than one year, having width 36 and 28 months respectively. The first break could be due to Donald Trump's election in the US; its CI includes almost the entire first presidency. Trump's election is significant as leaders across the board try to side or oppose his rhetoric and policies. The second break could be due to the COVID--19 pandemic which was a difficult multifaceted problem for leaders and the economy.

Table \ref{tab:EmpEx2:panel} presents some further robustness results examining how the number and location of breaks vary across model specifications. The parameters which vary are trimming, the constant/fixed effects specification, the significance level, and whether ``Elections" are included as non-breaking regressor. Overall, we observe that the panel results are robust to these choices.

\begin{table}[h]
\resizebox{\textwidth}{!}{ 
    \centering
    \begin{tabular}{l c c c c  c}\hline\hline
     & \multicolumn{3}{c}{Individual Fixed Effects} & \multicolumn{2}{c}{Overall Constant } \\
    Break \# & (1)              & (2)               & (3)                  & (4)             & (5)  \\ \hline
        1   &  2018m9           &  2018m9           &   2018m9           &  2018m9         &   2018m9 \\
            & (2017m3, 2020m3)  & (2017m3, 2020m3)  & (2017m3, 2020m3)   & (2017m3, 2020m3)& (2018m6, 2018m12) \\
        2   &  2020m4           & 2020m4            &  2020m4            &  2020m4         & 2020m4 \\
            & (2018m10, 2021m10)& (2018m10, 2021m10)& (2019m2,2021m6)    & (2019m2,2021m6) & (2020m2, 2020m6) \\
      \hline\hline        
UDmax & $15.60^{***}$ & $9.73^*$ & $13.91^{***}$ &  $13.85^{***}$  & $8.13^{***} $\\ \hline
        Break in constant & No & No & No  & No & Yes \\
        Non-breaking V.  & -   & -      & Election  &   Election & Election\\
        Trimming & 5\% & 5\%  & 5\%   & 5\%  & 5\% \\
        Critical V. & 1\% & 5\% & 1\%  & 1\% & 1 \% \\        
        \hline\hline
    \end{tabular}}
    \caption{\small 95\% Confidence intervals in parenthesis. UDmax is the test statistic of Hypothesis 2 with no breaks against \(1 < s < s_{max} = 19\). Stars indicate significant level at $^{***} 1\%$, $^{**} 5\%$, $^* 10\%$. Level Critical Values indicates the critical values at which the break dates are selected. 
    Non-breaking V. indicates non breaking variable.}
    \label{tab:EmpEx2:panel}
\end{table}

\begin{figure}
	\centering
	\includegraphics[width=\textwidth]{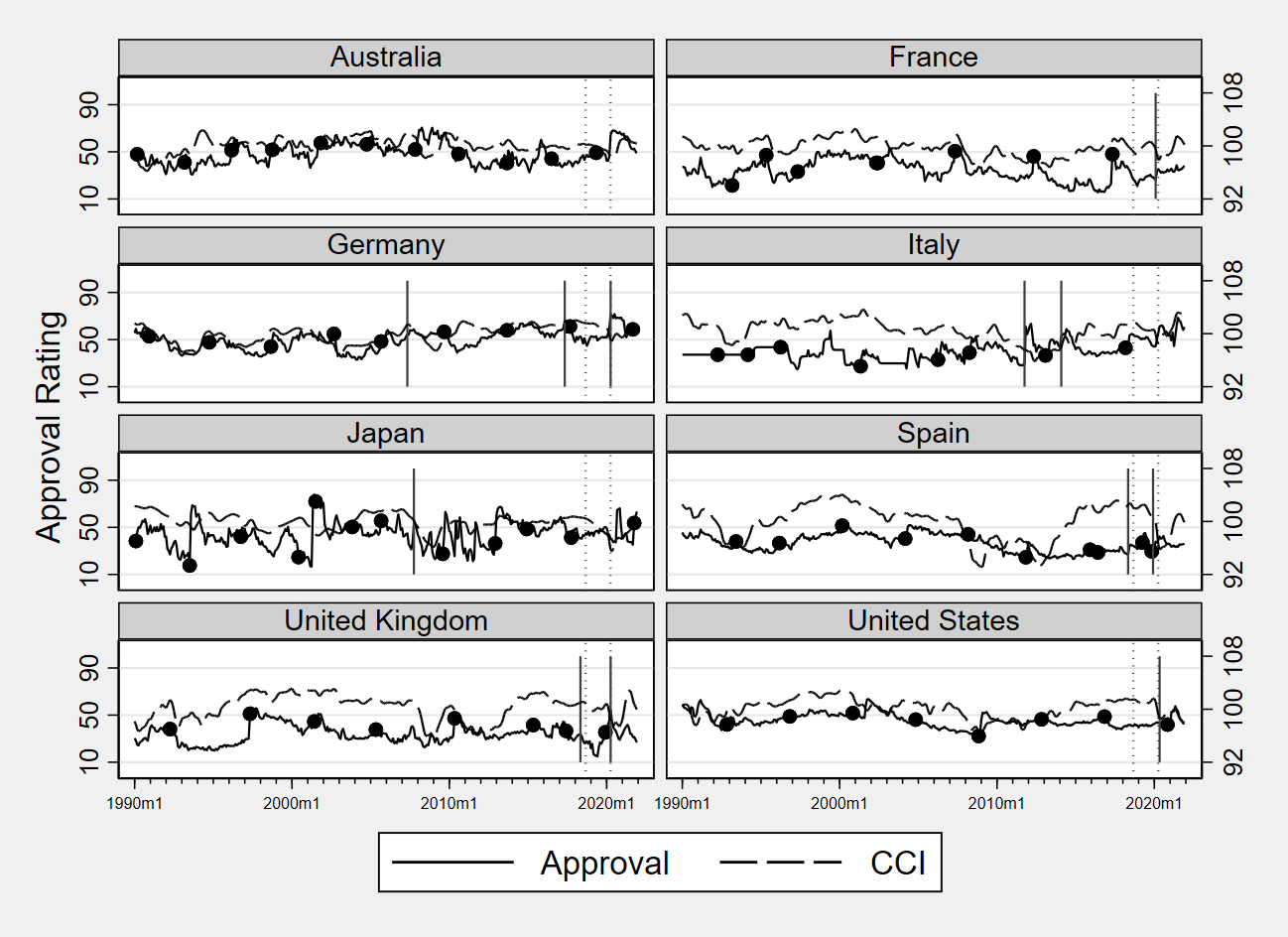}
	\caption{Leader Confidence. Dashed lines indicate break point estimates on country level, dotted on panel level and dots indicate elections.}
	\label{fig:Leaders}
\end{figure}

We now explore cross-country heterogeneity by analysing the disaggregated data for each of the 8 countries in the sample. Individual country analysis allows for heterogeneous intercept and slope regression coefficients, although it can be less efficient. The results are depicted in Figure \ref{fig:Leaders}, which includes the estimated break dates marked by vertical lines as they have been estimated in each time series. For comparison, the estimated breaks of the full panel are indicated by a dotted line. The number of estimated breaks ranges from none (Australia) to three (Germany). The April 2020 break estimated in the panel appears in five of the eight countries: France, Germany, Spain, the UK and the US. The break in 2018 appears only in Spain and the UK. Overall, the figure shows that the heterogeneity across countries is significant.

As a final exercise, we return to the panel case and assume that there is a single break. We now plot the SSRs \texttt{estat ssr} after using \texttt{xtbreak estimate}. The \texttt{estat ssr} function is only available after the estimation of a single break and displays the SSR for each possible break date.\footnote{The SSR function is different for different number of breaks. If there are two breaks, the SSR function becomes a surface.} 

\begin{stlog}
. xtbreak est d.approval d.CCI , trim(0.05) nobreakvar(ElectionQ) breaks(1) python
{\smallskip}
Estimation of break points
                                            Number of obs       =   3064
                                            Number of Groups    =      8
                                            Obs per group       =    383
                                            SSR                 =  34907.31
                                            Trimming            =      0.05
\HLI{80}
  \#      Index     Date                          [95\% Conf. Interval]
\HLI{80}
  1        360     2020m1                        2015m3         2024m11
\HLI{80}
{\smallskip}
. estat ssr , scheme(sj) name(ssr, replace)
{\smallskip}

\end{stlog}

\begin{figure}[!h]
	\centering
	\includegraphics[width=\textwidth]{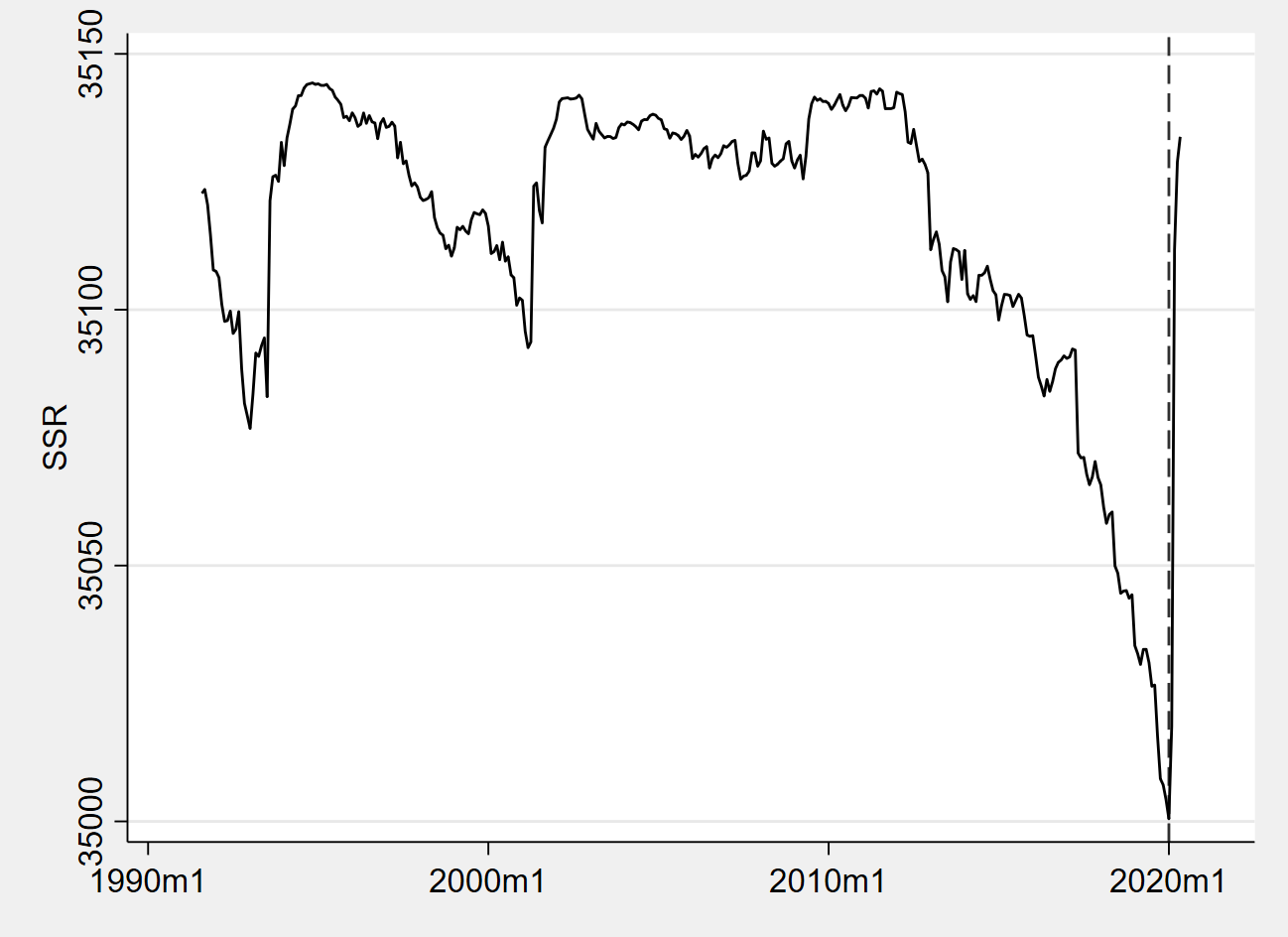}
	\caption{SSRs over time.}
	\label{fig:EX2_SSR}
\end{figure}

The estimated break point is indicated by a dashed line. The break is estimated to be in January 2020 and the SSR of 34907.31 represents the sum of the SSR of an estimation ranging from January 1990 until January 2020 and the SSR of an estimation from February 2020 to December 2021. This break is almost identical to the second break in the panel data model above, and coincides with the second break appearing in 5 out of the 8 countries of the sample. In an underspecified model in which the assumed number of breaks is less than the true number of breaks, the break date estimators are still consistent for the true break dates (see \cite{BaiPerron1998}). Therefore, the estimator of the one break here is meaningful and converges to one of the true break points; which of the true breaks it will converge to depends on the magnitudes of the breaks and the duration of the regimes. 

\section{Conclusion}

This paper presents a new community contributed command, called \texttt{xtbreak}, which enables researchers to detect breaks, and to estimate their number and location. The command can be applied to time series and panel data, and can hence be seen as a complete break detection toolbox that is applicable regardless of the structure of the data. In our empirical illustration, we employ US country-level and state-level data to investigate the relationship between the number of COVID--19 cases and deaths, which may well have changed as a result of improvements in testing capacity, reporting routines and treatments. While the time series data set suggests 2 breaks, the panel data set suggests 4. Moreover, 2 out of the 4 breaks detected using the panel data set coincide with those detected using the time series data set. The use of the relatively larger panel data therefore leads to the detection of two additional breaks which are not detected when using the time series data set. 

In a second empirical application we examine if there are breaks in the relationship between consumer confidence and the approval ratings of country leaders. In the panel of 8 countries we estimate two breaks, which however have wide confidence intervals. We analyse each country separately and find that there is great cross-country heterogeneity in terms of the number and locations of breaks. 

\section{How to install}

The latest version of the \texttt{xtbreak} package can be obtained by typing the following in Stata:

\begin{stlog}
	net from https://janditzen.github.io/xtbreak/
\end{stlog}

Updates and further documentation can be found on \href{https://janditzen.github.io/xtbreak/}{GitHub}.

\section{Acknowledgements}

We are grateful to all participants of the Swiss, German and US Stata User Group Meetings in 2020 and 2021. Ditzen acknowledges financial support from Italian Ministry MIUR under the PRIN project Hi-Di NET - Econometric Analysis of High Dimensional Models with Network Structures in Macroeconomics and Finance (grant 2017TA7TYC). Westerlund acknowledges financial support from the Knut and Alice Wallenberg Foundation through a Wallenberg Academy Fellowship.

\bibliographystyle{sj}
\bibliography{main}

\begin{aboutauthors}
Jan Ditzen is an Assistant Professor at the Faculty of Economics and Management of the Free University of Bozen-Bolzano, Italy. His research interests are in the field of applied econometrics and spatial econometrics, particularly cross-sectional dependence in large panels.
	
Yiannis Karavias is a Professor of Finance at Brunel University of London. His research focuses on panel data models and their applications in macroeconomics and finance. Methodological interests include structural change, threshold regression, non-stationarity, and optimal hypothesis testing.

Joakim Westerlund is a Professor at Lund University and at Deakin University. Joakim's primary research interest is the analysis of panel data. He has mainly been concerned with the development of procedures for estimation and testing of non-stationary panel data, but more recently his interest has shifted towards the development of econometric tools for factor-augmented regression models.
\end{aboutauthors}

\clearpage
\end{document}